\documentclass[11pt]{article}
\usepackage{amsmath,amsfonts,amsthm,amssymb,color}
\usepackage{mathrsfs}
\usepackage{fancybox}
\usepackage{framed,float,caption,algorithm2e}  
\usepackage{fullpage}
\usepackage{newclude}
\usepackage{hyperref}

\newtheorem{theorem}{Theorem}[section]
\newtheorem{corollary}[theorem]{Corollary}
\newtheorem{lemma}[theorem]{Lemma}
\newtheorem{observation}[theorem]{Observation}
\newtheorem{proposition}[theorem]{Proposition}

\newcommand{\Proofof}[1]{\smallskip\noindent\textit{\textbf{Proof of #1:}}\quad}

\theoremstyle{definition}
\newtheorem{definition}[theorem]{Definition}
\newtheorem{remark}[theorem]{Remark}

\newenvironment{fminipage}%
  {\begin{Sbox}\begin{minipage}}%
  {\end{minipage}\end{Sbox}\fbox{\TheSbox}}

\def\var#1{\mbox{\bf Var}\left[ #1 \right]}
\newcommand{\E}{\mbox{{\bf E}}}

\def\abs#1{\left|#1  \right|}

\def\norm#1{\left\| #1 \right\|}

\newcommand\R{\mathbb{R}}
\newcommand\rea{\mathbb{R}}

\newcommand\eps{\epsilon}

\newcommand {\ELSE}{{\bf else\ }}
\newcommand {\IF}{{\bf if\ }}

\newcommand {\RETURN}{\mbox{\bf Return\ }}

\DeclareMathOperator*{\med}{median}
\DeclareMathOperator*{\mean}{mean}
\DeclareMathOperator*{\pol}{poly}
\DeclareMathOperator*{\Tr}{Tr}

\newcommand\zero{\boldsymbol{0}}
\newcommand\pca{\textsc{AgnosticMean}}
\newcommand\outdamp{\textsc{OutlierDamping}}
\newcommand\outrunc{\textsc{OutlierTruncation}}
\newcommand\outrem{\textsc{OutlierRemoval}}
\newcommand\safeoutrunc{\textsc{SafeOutlierTruncation}}

\newcommand\covEst{\textsc{CovarianceEstimation}}
\newcommand\powit{\textsc{AgnosticOperatorNorm}}

\def\aa{\pmb{\mathit{a}}}
\newcommand\bb{\boldsymbol{\mathit{b}}}

\newcommand\ee{\boldsymbol{\mathit{e}}}

\newcommand\uu{\boldsymbol{\mathit{u}}}
\newcommand\vv{\boldsymbol{\mathit{v}}}
\newcommand\ww{\boldsymbol{\mathit{w}}}
\newcommand\yy{\boldsymbol{\mathit{y}}}
\newcommand\zz{\boldsymbol{\mathit{z}}}
\newcommand\xx{\boldsymbol{\mathit{x}}}

\newcommand\mmu{\boldsymbol{\mathit{\mu}}}
\newcommand\ddelta{\boldsymbol{\mathit{\delta}}}

\newcommand\sigmahat{{\widehat{\sigma}}}

\newcommand\SSigmac{\boldsymbol{{\Sigma^{(2)}}}}
\newcommand\SSigma{\boldsymbol{{\Sigma}}}
\renewcommand\AA{\boldsymbol{\mathit{A}}}
\newcommand\BB{\boldsymbol{\mathit{B}}}

\newcommand\II{\boldsymbol{\mathit{I}}}

\newcommand\PP{\boldsymbol{\mathit{P}}}

\newcommand\WW{\boldsymbol{\mathit{W}}}

\newcommand\YY{\boldsymbol{\mathit{Y}}}

\newcommand\dist{\mathcal{D}}

\newcommand\SSigmahat{\boldsymbol{\widehat{\Sigma}}}

\newcommand\mmuhat{\boldsymbol{\widehat{\mathit{\mu}}}}

\newcommand{\one}{\mathbf{1}}

\newcommand{\todo}[1]{}
\newcommand{\anup}[1]{}
\newcommand{\kevin}[1]{}
\newcommand{\santosh}[1]{}

\newcommand{\pr}[1]{\left(#1\right)}

\newcommand{\const}{C_4}

\title{Agnostic Estimation of Mean and Covariance
}

\author{Kevin A. Lai\thanks{Georgia Tech. Email: \{kevinlai, anup.rao, vempala\}@gatech.edu}
\and
Anup B. Rao\footnotemark[1]
\and
Santosh Vempala\footnotemark[1]
}

\begin{document}

\maketitle

\begin{abstract}
We consider the problem of estimating the mean and covariance of a distribution from iid samples in $\R^n$, in the presence of an $\eta$ fraction of malicious noise; this is in contrast to much recent work where the noise itself is assumed to be from a distribution of known type. The agnostic problem includes many interesting special cases, e.g., learning the parameters of a single Gaussian (or finding the best-fit Gaussian) when $\eta$ fraction of data is adversarially corrupted, agnostically learning a mixture of Gaussians, agnostic ICA, etc. We present polynomial-time algorithms to estimate the mean and covariance with error guarantees in terms of  information-theoretic lower bounds. As a corollary, we also obtain an agnostic algorithm for Singular Value Decomposition.
\end{abstract}

\thispagestyle{empty}
\newpage

\setcounter{page}{1}

\section{Introduction}
 
The mean and covariance of a probability distribution are its most basic parameters (if they are bounded). Many families of distributions are defined using only these parameters. Estimating the mean and covariance from iid samples is thus a fundamental and classical problem in statistics. The sample mean and sample covariance are generally the best possible estimators (under mild conditions on the distribution such as their existence). However, they are highly sensitive to noise. The main goal of this paper is to estimate the mean, covariance and related functions in spite of arbitrary (adversarial) noise.

Methods for efficient estimation, in terms of sample complexity and time complexity, play an important role in many algorithms.
One such class of problems is unsupervised learning of generative models. Here the input data is assumed to be iid from an unknown distribution of a known type. The classical instantiation is Gaussian mixture models, but many other models have been studied widely. These include topic models, stochastic block models, Independent Component Analysis (ICA) etc. In all these cases, the problem is to estimate the parameters of the underlying distribution from samples. For example, for a mixture of $k$ Gaussians in $\R^n$, it is known that the sample and time complexity are bounded by $n^{O(k)}$ in general \cite{Kalai2010, Moitra2010, Belkin2010} and by poly($n,k$) under natural separation assumptions \cite{dasgupta1999learning, sanjeev2001learning,  vempala2004spectral, dasgupta2007probabilistic, chaudhuri2008learning, brubaker2008isotropic, HsuK13}. For ICA, samples are of the form $Ax$ where $A$ is unknown and $x$ is chosen randomly from an unknown (non-Gaussian) product distribution; the problem is to estimate the linear transformation $A$ and thus unravel the underlying product structure \cite{FriezeJK96, NguyenR09,  Cardoso1998multidimensional,ICAbook, ComonJutten, BelkinRV12, AroraGMS12,  bhaskara2013uniqueness, goyal2014fourier, VempalaX15}. 
These, and other models (see e.g., \cite{kannan2009spectral}), have been a rich and active subject of study in recent years and have lead to interesting algorithms and analyses. 
 
The Achilles heel of algorithms for generative models is the assumption that data is {\em exactly} from the model. This is crucial for known guarantees, and relaxations of it are few and specialized, e.g., in ICA,  data could by noisy, but the noise itself is assumed to be Gaussian. Assumptions about rank and sparsity are made in a technique that is now called Robust PCA \cite{Chandrasekaran2011,Candes2011,Xu2010}. There have been attempts \cite{Kwak2008,Mccoy2011} at achieving robustness by L1 minimization, but they don't give any error bounds on the output produced.  A natural, important and wide open problem is estimating the parameters of generative models in the presence of arbitrary, i.e., {\em malicious} noise, a setting usually referred to as {\em agnostic} learning. The simplest version of this problem is to {\em estimate a single Gaussian} in the presence of malicious noise. Alternatively, this can be posed as the problem of finding a best-fit Gaussian to data or agnostically learning a single Gaussian. We consider the following generalization:

\paragraph{Problem 1 [Mean and Covariance]} {\em Given points in $\R^n$ that are each, with probability $1-\eta$ from an unknown distribution with mean $\mu$ and covariance $\Sigma$, and with probability $\eta$ completely arbitrary, estimate $\mu$ and $\Sigma$.}\\

There is a large literature on {\em robust} statistics (see e.g., \cite{Huber2011,Hampel2011,Maronna2006}), with the goal of finding estimators that are stable under perturbations of the data. The classic example for points on a line is that the sample median is a robust estimator while the sample mean is not (a single data point can change the mean arbitrarily). One measure for robustness of an estimator is called {\it breakdown } point, which is the minimum fraction of noise that can make the estimator arbitrarily bad.  Robust statistics have been proposed and studied for mean and covariance estimation in high dimension as well (see~\cite{Huber1964,Tukey74, Maronna1976, Devlin1981, Donoho1982,Davies1987,Lopuhaa1991, Donoho1992, Maronna1992,Maronna2012, Chen2015} and the references therein). Most commonly used methods (including M-estimators) to estimate the covariance matrix were shown to have very low break down points \cite{Donoho1982}. The notion of robustness we consider quantifies how far the estimated value is from the true value.  To the best of our knowledge,  all the papers either suffer from the difficulty that their algorithms are computationally very expensive, namely exponential time in the dimension, or have poor or no guarantees for the output. Tukey's median \cite{Tukey74}) is an example of the former. It is defined as the {\it deepest } point with respect to a given set of points $\{ \xx_i \}_i.$ As proven in \cite{Chen2015}, this is an optimal estimate of the mean. But there is no known polynomial time algorithm to compute this.  Another well-known proposal (see \cite{Small90}) is the geometric median:
$$\arg \min_{\yy } \sum_{i} \|\yy - \xx_i \|_2.$$
 This has the advantage that it can be computed via a convex program. Unfortunately, as we observe here (see Proposition~\ref{prp:geomMed}), the error of the mean  estimate produced by this method grows polynomially  with the dimension (also see \cite{Bruce11}). 
 
This leads to the question, what is the best approximation one can hope for with $\eta$ arbitrary (adversarial) noise. From a purely information-theoretic point of view, it is not hard to see that even for a single Gaussian $N(\mu, \sigma^2)$  in one dimension, the best possible estimation of the mean will have error as large as $\Omega(\eta\sigma)$, i.e., any estimate $\tilde{\mu}$ can be forced to have $\|\mu -\tilde{\mu}\| = \Omega(\eta \sigma)$.  For a more general distribution, this can be slightly worse, namely, $\Omega(\eta^{3/4}\sigma)$ (see Section \ref{sec:lowerBounds}). What about in $\R^n$? Perhaps surprisingly, but without much difficulty, one can show that the information-theoretic upper bound matches the lower bound in any dimension, with no dependence on the dimension. This raises a compelling algorithmic question: what are the best estimates for the mean and covariance that can be computed efficiently? 

In this paper, we give polynomial time algorithms to estimate the mean with error that is close to the information-theoretically optimal estimator. The dependence on the dimension, of the error in the estimated mean, is only $\sqrt{\log n}$. To the best of our knowledge, this is the first polynomial-time algorithm with an error dependence on dimension that is less than $\sqrt{n}$, the bound achieved by the geometric median. Moreover, as we state precisely later, our techniques extend to very general input distributions and to estimating higher moments.

 Our algorithm is practical. A matlab implementation for mean estimation can be found in \cite{KRV16}. It takes less a couple of seconds to run on a $500$-dimensional problem with $5000$ samples on a personal laptop. 

\paragraph{Model.} 
We are given points $\xx_1,...,\xx_m \in \rea^n$ sampled according to the following rule.  With $1-\eta$ probability each $\xx_i$ is independently sampled from a distribution $\dist$ with mean $\mmu$ and covariance $\SSigma$,  and with $\eta$ probability it is picked by an adversary.  For ease of notation, we will write $\xx_i \sim \dist_{\eta}$ when we want to say the $\xx_i$ is picked according to the above rule. The problem we are interested in is to estimate $\mmu$ and $\SSigma$ given the samples. In the following, we will consider mainly two kinds of distributions. 
\begin{description}
\item[Gaussian]  $\dist = N(\mmu, \SSigma)$ is the Gaussian with mean $\mmu$ and covariance $\SSigma$.
\item[Bounded Moments] Let $\dist$ is a distribution with mean $\mmu$ and covariance $\SSigma$. We say it has  bounded $2k$'th
moments if there exists a constant $C_{2k}$ such that for every unit vector $\vv$, 
\begin{equation}
\label{eq:boundedMom}
\E \left((\xx-\mmu)^T \vv\right)^{2k} \leq  C_{2k} \left(\E \left((\xx - \mmu)^T \vv \right)^2 \right )^{k}=C_{2k} (\var{\xx^T \vv})^k.
\end{equation}
Here $\var{\xx^T \vv} = \left( \vv^T\SSigma\vv \right)^2$ is the variance of $\xx$ along $\vv.$ For mean estimation, $C_4$ will be used, and for covariance estimation, $C_8$ will be needed.
\end{description}

\subsection{Main results}

All the results we state hold with probability $1 - 1/\pol(n)$ unless otherwise mentioned. We will also assume $\eta$ is a less than a universal constant. We begin with agnostic mean estimation.

\begin{theorem}[Gaussian mean]
\label{thm:mainGauss}
Let
$\dist =  N(\mmu, \SSigma)$, $\mmu \in \rea^n$. 
There exists a $\pol(n,1/\eps)$-time algorithm that takes as input  $m = O\pr{ \frac{n (\log n + \log 1/\eps) \log n}{\eps^2} }$ independent samples $\xx_1,...,\xx_m \sim \dist_{\eta}$ and computes $\mmuhat$  such that the error 
$\| \mmu - \mmuhat \| _2$ is bounded as follows:
\[
\begin{array}{cc}
O\left( \eta + \eps \right)  \sigma  \sqrt{ \log n} & \text{ if }  \SSigma = \sigma^2 \II  \\
O \left( \eta^{1/2} + \eps  \right) \| \SSigma \|^{1/2}_2 \log^{1/2} n & \text{ otherwise. }
\end{array} 
\]
\end{theorem}
We note that the sample complexity is nearly linear, and almost matches the complexity for mean estimation with no noise. 

\begin{remark}
If we take $m = O\pr{ \frac{n^2 (\log n + \log 1/\eta) \log n}{\eta^2} }$ samples, and assume that $\eta < c/\log n$ for a small enough constant $c > 0$, then by combining theorems~\ref{thm:mainCovEst} and \ref{thm:mainGauss}, we can improve the $\eta$ dependence for the non-spherical Gaussian case in Theorem~\ref{thm:mainGauss} to $\| \mmu - \mmuhat \| _2 = O \left( \eta^{3/4} \right)  \| \SSigma \|^{1/2}_2 \log^{1/2} n .$ 
\end{remark}
Our next theorem is a similar result for much more general distributions. 

\begin{theorem}[General mean]
\label{thm:mainGeneral}
Let $\dist$ be a distribution on $\rea^n$ with mean $\mmu$, covariance $\SSigma$, and bounded fourth moments (see Equation~\ref{eq:boundedMom}). There exists a $\pol(n,1/\eps)$-time algorithm  that takes as input a parameter $\eta$  and $m  = O\pr{ \frac{n (\log n + \log 1/\eps) \log n}{\eps^2} }$ independent samples $\xx_1,...,\xx_m \sim \dist_{\eta},$ and computes  $\mmuhat$  such that 
 the error $\| \mmu - \mmuhat \| _2$ is bounded as follows:
\[
\begin{array}{cc}
O\left(  \const^{1/4}  (\eta + \eps)^{3/4}  \right)  \sigma \sqrt{ \log n} & \text{ if }  \SSigma = \sigma^2 \II  \\
O  \pr{\eta^{1/2} + \const^{1/4} (\eta + \eps)^{3/4} }\| \SSigma \|^{1/2}_2 \log^{1/2} n  & \text{ otherwise. }
\end{array} 
\]
\end{theorem}


The bounds above are nearly the best possible (up to a factor of $O(\sqrt{\log n})$) when the covariance is a multiple of the identity. 
\begin{observation}[Lower Bounds]
\label{thm:mainLower}
Let $\dist$ be a distribution with mean $\mmu \in \rea^n$ and covariance $\SSigma$. Any algorithm that takes $m$ 
(not necessarily $O(\pol(n))$)  samples $\xx_1,...,\xx_m \sim \dist_{\eta},$  and computes a $\mmuhat$ should have with constant probability
 the error $\| \mmu - \mmuhat \| _2$ is  
\[
\begin{array}{cc}
\Omega(\eta \sqrt{ \| \SSigma \| _2}) & \text{ if } \dist = N(\mmu,\SSigma) \\
\Omega(\eta^{3/4} \sqrt{ \| \SSigma \| _2}) & \text{ if } \dist \text{ has bounded fourth moments.}
\end{array}
\] 
\end{observation}

\begin{theorem}[Covariance Estimation]
\label{thm:mainCovEst}
Let $\dist$ be a distribution with mean $\mmu$ and covariance $\SSigma$ and that 
(a) for $\xx \sim \dist$, $\xx$ and $(\xx - \mmu) (\xx - \mmu)^T $ have bounded fourth moments with constants $\const$ and $C_{4,2}$(see Equation~\ref{eq:boundedMom}) respectively. 
(b) $\dist$ is an (unknown) affine transformation of a $4$-wise independent distribution. Then, there is an algorithm that takes as input $m  = O\pr{ \frac{n^2 (\log n + \log 1/\eps) \log n}{\eps^2} }$ samples $\xx_1,...\xx_m \sim \dist_{\eta}$ and $\eta$ and computes in $\pol(n,1/\eps)$-time a covariance estimate $\SSigmahat$ such that
$$\|  \SSigmahat - \SSigma \| _F =  O  \pr{\eta^{1/2} + C_{4,2}^{1/4} (\eta + \eps)^{3/4} }\const^{1/2} \| \SSigma \|_2 \log^{1/2} n$$
where $ \|  \cdot  \| _F$ denotes the Frobenius norm.
\end{theorem}
If $\dist = N(\mmu, \SSigma),$ then it satisfies the hypothesis of the above theorem. More generally, it holds for any $8$-wise independent distribution with bounded eighth moments and whose fourth moment along any direction is at least $(1+c)$ times the square of the second moment for some $c >0$. We also note that if the distribution is isotropic, then covariance estimation is essentially a $1$-d problem and we get a better bound.


\begin{theorem}[Agnostic 2-norm]
\label{thm:main2norm}
Suppose $\dist$ is a distribution which satisfies the following concentration inequality: there exists a constant $\gamma$ such that for every unit vector $\vv$
$$\Pr \left( \left |(\xx - \mmu)^T \vv \right| > t \sqrt{\vv^T\SSigma\vv} \right ) \leq e^{-t^{\gamma}}.$$
 Then, there is an algorithm that runs in $\pol(n,\log \frac{1}{\eta})$ time that takes as input $\eta$ and $m= O \pr{ \frac{n^3 (\log n/\eta)^2 \log n}{\eta^2} } $ independent samples $\xx_1,...,\xx_m \sim \dist_{\eta}$, and computes $\widehat{\lambda}_{\max}$ such that
$$ \left( 1 - O(\eta) \right)  \| \SSigma \| _2 \leq \widehat{\lambda}_{\max}  \leq \left( 1 + O(\eta \log^{2/\gamma} n/\eta) \right)  \| \SSigma \| _2 .$$ 
\end{theorem}
In independent work, \cite{DKKL16} gave a similar algorithm, which they call a Gaussian filtering method, for agnostic mean estimation assuming a spherical covariance matrix; while their guarantees are specifically for Gaussians, the error term in their guarantee grows only with $\log(1/\eta)$ rather than $\log n$. They also give a completely different algorithm based on the Ellipsoid method, for a simple family of distributions including Gaussian and Bernoulli. 

As a corollary of Theorem~\ref{thm:mainCovEst}, we get a guarantee for agnostic SVD. 
\begin{theorem}[Agnostic SVD]
\label{thm:agnosticSVD}
Let $\dist$ is a distribution that satisfies the hypothesis of Theorem~\ref{thm:mainCovEst}. Let $\SSigma_k$ be the best rank $k$ approximation to $\SSigma$ in $\| \cdot \|_F$ norm. There exists a polynomial time algorithm that takes as input $\eta$ and $m = \pol(n)$ samples from $\dist_{\eta}.$ It produces a rank $k$ matrix $\SSigmahat_k$ such that 
$$ \left \| \SSigma - \SSigmahat_k \right \|_F \leq \left \| \SSigma - \SSigma_k \right \|_F + 
O \left( \sqrt{\eta \log n  }\right)  \| \SSigma \|_2 .$$
\end{theorem}

Given the wide applicability of SVD to data, we expect the above theorem will have many applications. As an illustration, we derive a guarantee for agnostic Independent Component Analysis (ICA). In standard ICA, input data points $x$ are generated as $As$ with a fixed unknown $n \times n$ full-rank matrix $A$ and $s$ generated from an unknown product distribution with non-Gaussian components. The problem is to estimate the matrix $A$ (the ``basis") from a polynomial number of samples in polytime. There is a large literature of algorithms for this problem and its extensions \cite{FriezeJK96, NguyenR09,  Cardoso1998multidimensional,ICAbook, ComonJutten, BelkinRV12, AroraGMS12,  bhaskara2013uniqueness, goyal2014fourier}. However, all these algorithms rely on no noise or the noise being random (typically Gaussian) and require estimating singular values to within $1/\pol(n)$ accuracy, and therefore unable to handle adversarial noise. On the other hand, the algorithm from \cite{VempalaX15}, which gives a sample complexity of $\tilde{O}(n)$, only requires estimating singular values to within $1/\pol(\log n)$. Our algorithm for agnostic SVD together with the Recursive Fourier PCA algorithm of \cite{VempalaX15} results in an efficient algorithm for agnostic ICA, tolerating noise $\eta = O(1/\log^c n)$ for a fixed constant $c$. To the best of our knowledge, this is the first polynomial-time algorithm that can handle more than an inverse $\pol(n)$ amount of noise.

\begin{theorem}[Agnostic Standard ICA]
\label{thm:agnosticICA}
  Let $x \in \R^n$ be given by a noisy ICA model $x=As$ with probability $1-\eta$ and be arbitrary with probability $\eta$, where $A \in \R^{ n
    \times n}$ has condition number $\kappa$, the components of $s$ are independent, $\norm{s} \le K
  \sqrt{n}$ almost surely, and for each $i$, $\E{s_i}=0, \E{s_i^2}=1$, $|\E{|s_i|^4}-3|
  \ge \Delta$ and $\max_i \E{|s_i|^5}\le M$. Then for any $\eps < \Delta^3/(10^8 M^2 \log^3 n), 1/(\kappa^4\log n)$ and
$\eta < \eps/2$, there is an algorithm that, with high probability, 
 finds vectors $\{b_1, \ldots, b_n\}$
  such that there exist signs $\xi_i = \pm 1$ satisfying 
$\norm{A^{(i)} - \xi_i b_i} \le \eps\|A\|_2$
  for each column $A^{(i)}$ of $A$, using $\pol(n,K,\Delta,M,\kappa, \frac{1}{\eps})$ 
samples. The running time is bounded by the time to compute $\tilde{O}(n)$ SVDs on real symmetric matrices of size $n \times n$.
\end{theorem}

Our results can also be used to estimate the mean and covariance of noisy Bernoulli product distributions, i.e. distributions in which each coordinate $i$ is 1 with probability $p_i$ and 0 with probability $1-p_i$. In one dimension, $\const$ for a Bernoulli distribution is $\frac{(1-p)^2}{p}+\frac{p^2}{1-p}$. For a Bernoulli product distribution, $\const$ will be within a constant of $\max_i \left\{ \frac{(1-p_i)^2}{p_i}+\frac{p_i^2}{1-p_i} \right\}$. Then Theorem \ref{thm:mainGeneral} can be applied to get an estimate $\mmuhat$ for the mean. For instance, if $\forall i, p_i = p$ and $p \ge \frac{1}{2}$, then $\|\mmu -\mmuhat\|_2 = O\pr{\sqrt{\eta(1+\sqrt{\eta p}) p \log n }}$. If $\const$ is constant, then by  Theorem \ref{thm:mainCovEst}, we can get an estimate for the covariance.


\section{Main Ideas}
\label{sec:mainIdeas}
Here we discuss the key ideas of the algorithms.
The algorithm $\pca$ (Algorithm~\ref{fig:pca}) alternates between an outlier removal step and projection onto the top $n/2$ principal components; these steps are repeated. It is inspired by the work of Brubaker \cite{Brubaker2009} who gave an agnostic algorithm for learning a mixture of well-separated spherical Gaussians.

For illustration, let us assume for now that the underlying distribution is $\dist =  N(\mmu, \sigma^2 \II).$ We are  given a set $S$ of  $m = \pol(n)$ points from $\dist_{\eta}$, and $S = S_G \cup S_N$ be the points sampled from the Gaussian and the adversary respectively. Let us also assume that $|S_N| = \eta |S|$. We will use the notation $\mmu_T$ for mean of the points in a set $T$, and $\SSigma_T$ for covariance of the points in $T$. We then have 
\begin{equation}
\label{eq:covSum}
\SSigma_{S} = (1-\eta) \sigma^2 \II + \eta  \SSigma_{S_N} + \eta(1-\eta)(\mmu_S - \mmu_N)(\mmu_S - \mmu_N)^T.
\end{equation}
 If the dimension is $n=1$, then we can show that the median of $S$ is an estimate for $\mmu$ correct up to an additive error of $O(\eta \sigma).$  Even if we just knew the direction of the  {\it mean shift} $ \mmu_S - \mmu = \eta (\mmu_G - \mmu_N)$, then we can estimate $\mmu$ by first projecting the sample $S$ on the line along $\mmu - \mmu_S$ and then finding the median. This would give an estimator $\mmuhat$ satisfying $\| \mmuhat - \mmu \|_2 = O(\eta \sigma).$ So we can focus on finding the direction of $\mmu _S- \mmu$. One would guess that the top principal component of the covariance matrix of $S$ would be a good candidate. But it is easy for the adversary to choose $S_N$ to make this completely useless.  
Since the noise points $S_N$ can be anything, just two points from $S_N$ placed far away on either side of the mean $\mmu$ along a particular line passing through $\mmu$ are sufficient to make the variance in that direction blow up arbitrarily. But we can limit this effect to some extent by an outlier removal step. 
By a standard concentration inequality for Gaussians, we know that the points in $S_G$ lie in a ball of radius $O(\sigma \sqrt{n})$ around the mean. So, if we can just find a point inside or close to the convex hull of the Gaussian and throw away all the points that lie outside a ball of  radius  $C \sigma \sqrt{n}$ around this point, we preserve all the points in $S_G$. This will also contain the effect of noise points on the variance since now they are restricted to be within $O(\sigma \sqrt{n})$ distance of $\mmu.$ We will see later that we can use coordinate-wise median as the center of the ball. By computing the variance by projecting onto any direction, we can figure out $\sigma^2$ up to a $1 \pm O(\eta)$ factor. From now on, we assume that all points in $S$ lie within a ball of radius $O(\sigma \sqrt{n})$ centered at $\mmu.$ 

 But even after this restriction, the top principal component may not contain any information about the mean shift direction. By just placing (say) $\eta/10$ noise points along the $e_1$ direction at $\pm \sigma \sqrt{n}$, and all the remaining noise points perpendicular to this at a single point at a smaller distance, we can make $e_1$ the top principal component. But $e_1$ is perpendicular to the mean shift direction.

 The idea to get around this is that even if the top principal component of $\SSigma_S$ may not be along the mean-shift direction, the span (call it $V$) of top $n/2$ principal components of $\SSigma_S$ will contain a big projection of the mean-shift vector. This is because, if a big component of the the mean-shift vector was in the span (say $W$) of bottom $n/2$ principal components  of $\SSigma_S$, by Equation~\ref{eq:covSum} this would mean that there is a vector  in $W$ with a large Rayleigh quotient. This implies that the top $n/2$ eigenvalues of $\SSigma_S$ are all big. Since $\SSigma_S = (1-\eta) \sigma^2 \II + \AA$, where $\AA = \eta  \SSigma_{S_N} + \eta(1-\eta)(\mmu_S - \mmu_N)(\mmu_S - \mmu_N)^T$, this is possible only if  $\Tr(\AA)$ is large.  But since the distance of each point in $S$ from $\mmu$ is $O(\sigma \sqrt{n})$, the trace of  $\AA$ cannot be too large. 
Therefore, in the space $W$, we can just compute the sample mean $\PP_W \mmu_S$ and it will be close to $\PP_W \mmu$.
We still have to find the mean in the space $V$. But we do this by recursing the above procedure in $V$. At the end we will be left with a one-dimensional space, and then we can just find the median. This recursive projection onto the top $n/2$ principal components is done in  Algorithm~\ref{fig:pca} . 

This generalizes to the non-spherical Gaussians with a few modifications. We use a different outlier removal step. In the non-spherical case, it is not trivial to compute $\|\SSigma\|_2$ to be used as the radius of the ball. We give an algorithm for this later on. To limit the effect of noise, we use a damping function. Instead of discarding points outside a certain radius, we damp every point by a weight so that further away points get lower weights. This is done in $\outdamp$ (Algorithm~\ref{fig:OutlierDamping}).  We get the guarantees of Theorem~\ref{thm:mainGauss} by running $\pca$ (Algorithm~\ref{fig:pca}) with the outlier removal routine being $\outdamp$. A detailed proof of the whole algorithm is given in Section~\ref{sec:mainGauss}.

We then turn to more general distributions which have bounded fourth moments. We need bounded fourth moments to ensure that the mean and covariance matrix of the distribution $\dist$ do not change much even after conditioning by an event that occurs with probability $1-\eta$. One difficulty for general distributions is that the outlier damping doesn't work. So for distributions $\dist$ with bounded fourth moments, we have another outlier removal routine called $\outrunc(\cdot,\eta).$ In this routine, we first find a point analogous to the coordinate-wise median for the Gaussians, and then consider a ball big enough to contain $1-\eta$ fraction of $S$. We throw away all the points outside this ball.  We get the guarantees of Theorem~\ref{thm:mainGeneral} by running $\pca$ (Algorithm~\ref{fig:pca}) with the outlier removal routine being $\outrunc$ (Algorithm~\ref{fig:OutlierTruncation}). The complete proof of this appears in Section~\ref{sec:mainGeneral}. 

We now have an algorithm to estimate the mean of very general (with bounded fourth moments) distributions. To estimate the covariance matrix, we observed that the covariance matrix of a distribution $\dist$ is given by $\E_{\dist} (\xx - \mmu)(\xx -\mmu)^T.$ If we knew what $\mmu$ was, then covariance can be computed by estimating the mean of the second moments. To compute the mean of the second moments, we can treat $(\xx - \mmu)(\xx -\mmu)^T$ as a vector in $n^2$ dimensions and run the algorithm for mean estimation. Also, we can estimate $\mmu$ by the same algorithm. Therefore,  we get  Theorem~\ref{thm:mainCovEst} by running $\covEst$  (Algorithm~\ref{fig:covEst}). Its proof appears in Section~\ref{sec:mainCovEst}.  

Algorithm $\powit$ (Algorithm~\ref{fig:powit}) estimates the $2$-norm $\| \SSigma \|_2$ for general distributions.
For illustration, suppose $\dist = N(\mmu, \SSigma),$ and we are given $m=\pol(n)$ samples $\xx_1,...,\xx_m\sim\dist_{\eta},$ and the mean $\mmu$.  We  consider the covariance-like matrix 
$$\SSigma(S,\mmu) = \frac{1}{m} \sum_i (\xx_i - \mmu)(\xx_i -\mmu)^T.$$ Since $1-\eta$ fraction of the points in $S$ are from the Gaussian, we have $\SSigma(S,\mmu) \succeq (1-\eta) \SSigma.$ Therefore, the top eigenvalue $\sigma^2$ of $\SSigma(S,\mmu)$ is at least $(1-\eta) \|\SSigma \|_2.$ Let $\vv$ be the top eigenvector of $\SSigma(S,\mmu).$ If the Gaussian variance along $\vv$ (which can be computed up to $1 \pm \eta$ factor) is much less than $\sigma^2$, this should be because there are a lot of noise points in $S$ whose projections onto $\vv$ are big compared to the projection of Gaussian points in $S$. We remove points in $S$ that have big projection and then iterate the entire procedure. We later show that this procedure terminates in $\pol(n)$ steps and when it terminates the top eigenvalue of $\SSigma(S,\mmu)$ is close to that of $\SSigma.$ A proof of this appears in Section~\ref{sec:main2norm}.

Theorem~\ref{thm:agnosticSVD} follows easily from Theorem~\ref{thm:mainCovEst}. Let $\SSigmahat_k$ be the top-$k$ eigenspace of $\SSigmahat$ from Theorem~~\ref{thm:mainCovEst}. We then have
\begin{align*}
\left \| \SSigma - \SSigmahat_k \right \|_F & \stackrel{(a)}{\leq}  \left \| \SSigma - \SSigmahat \right \|_F +  \left \| \SSigmahat - \SSigmahat_k \right \|_F \\
&\stackrel{(b)}{\leq} \left \| \SSigma - \SSigmahat \right \|_F + \left \| \SSigmahat - \SSigma_k \right \|_F \\
& \stackrel{(c)}{\leq} 2 \left  \| \SSigma - \SSigmahat \right \|_F + \left \| \SSigma - \SSigma_k \right \|_F \\
& \stackrel{(d)}{\leq} \left \| \SSigma - \SSigma_k \right \|_F + O \left( \sqrt{\eta \log n  }\right)  \| \SSigma \|_2 .
\end{align*}
$(a),(c)$ follow from triangle inequality, $(b)$ follows from the fact that $\SSigmahat_k$ is the best rank-$k$ approximation and $(d)$  from the guarantees of Theorem~\ref{thm:mainCovEst}.

Finally we outline the application to agnostic ICA. The algorithm from \cite{VempalaX15}. Proceeds by first estimating the mean and covariance, in order to make the underlying distribution isotropic. Here we estimate the covariance matrix $\SSigma$ by $\hat{\SSigma}$ and use it to determine a new isotropic transformation $\hat{\SSigma}^{-\frac{1}{2}}$. Since our agnostic SVD algorithm gives a guarantee of $\|\SSigma - \tilde{\SSigma}\|_F \le O(\sqrt{\nu \log n})\|\SSigma\|_2$, the isotropic transformation results in a guarantee of
\[
\|\hat{\SSigma}^{-\frac{1}{2}}\SSigma\hat{\SSigma}^{-\frac{1}{2}}-I\|_2 \le O(\sqrt{\eta\log n})\frac{\|\SSigma\|_2}{\|\SSigma^{-1}\|_2} = O(\sqrt{\eta \log n}\kappa^2).
\]
Next the algorithm estimates a weighted covariance matrix $\WW$ with the weight of a point $\xx$ proportional to $\cos(\uu^T\xx)$ for $\uu$ chosen from a Gaussian distribution; it computes the SVD of $\WW$. For this we use our algorithm again (the weights are applied individually to each sample). The main guarantee is that the eigenvectors of this weighted covariance approximate the columns of $A$. This relies on the maximum eigenvalue gap of $\WW$ being large, and it has to be approximated to within additive error $\eps = O(1/(\log n)^3)$. Theorem \ref{thm:agnosticSVD} implies that the additional error in eigenvalues is bounded by 
$O(\sqrt{\eta\log n})\|\SSigma\|_2$, and therefore it suffices to have $\sqrt{\eta\log n} < c/(\log n)^3$ for a sufficiently small constant $c$ that depends only on the cumulant and moment bound assumptions (i.e., $\Delta, M$). Thus, if suffices to have $\eta < \eps/2 \le c(\log n)^{-7}$.   

\subsection{Lower Bounds: Observation~\ref{thm:mainLower}}
\label{sec:lowerBounds}
In this section we will show  the lower bounds stated in Observation~\ref{thm:mainLower}. For Gaussian distributions, this is a special case of a theorem proved in~\cite{Chen2015}. We reproduce the relevant part here for completeness. We will show that there are distributions $\dist_1 = N(\mmu_1,\sigma^2 \II), \dist_2 = N(\mmu_2,\sigma^2 \II)$ and distributions $Q_1,Q_2$ such that $\| \mmu_1 - \mmu_2 \|_2 = \Omega(\eta \sigma)$ and 
\begin{equation}
\label{eq:identityGauss}
\dist_{\eta} = (1-\eta)\dist_1 + \eta Q_1 = (1-\eta)\dist_2 + \eta Q_2.
\end{equation}
So, given $\dist_{\eta}$, no algorithm can distinguish between $\dist_1,\dist_2$. Let $\phi_1$ be p.d.f of $\dist_1$ and $\phi_2$ be the p.d.f of $\dist_2.$ Let $\mmu_1,\mmu_2$ be such that the total variation distance between $\dist_1, \dist_2$ is
$$\frac{1}{2}\int |\phi_1 - \phi_2| dx = \frac{\eta}{1-\eta}.$$    
By a standard inequality for the total variation distance of Gaussian distributions, this implies that $\| \mmu_1 - \mmu_2 \|_2 \geq \frac{2\eta \sigma}{1-\eta}.$ Let $Q_1$ be the distribution with p.d.f $\frac{1-\eta}{\eta} (\phi_2 - \phi_1) \one_{\phi_2 \geq \phi_1}$ and $Q_2$ be the distribution with p.d.f $ \frac{1-\eta}{\eta}(\phi_1 - \phi_2) \one_{\phi_1 \geq \phi_2}$. It is now easy to verify that Equation~\ref{eq:identityGauss} is satisfied. This proves item one of Observation~\ref{thm:mainLower}.

For the distributions with bounded fourth moments, consider the following two one-dimensional distributions. $\dist_1$ is supported on two points $\{ -\sigma, \sigma \}$ with the corresponding probabilities $\{ 1/2, 1/2 \}$. $\dist_2$ is supported on three points $\{ -\sigma, \sigma, \sigma/\eta^{1/4} \}$ with probabilities $\{ (1-\eta)/2, (1-\eta)/2, \eta \}$ respectively. Let   $\eta \leq 1/4$. It is easy to check that both $\dist_1$ and $\dist_2$ have bounded fourth moments with the constant $\const = 8$. Furthermore, $\dist_2$ can be obtained from $\dist_1$ by adding $\eta$ fraction of noise points. So no algorithm can distinguish between the two distributions. Since their means differ by $\eta^{3/4} \sigma$, no algorithm can get an estimate better than this.

We will now show that the geometric median:
$$\arg \min_{\yy } \sum_{i} \| \xx_i - \yy \|_2$$
has a $\sqrt{n}$ dependence on the dimension. We show this in the Gaussian case even if we have access to the whole distribution, but with  $\eta$ fraction of noise points placed all at a single point far away from most of the Gaussian points. 

\begin{proposition}[Geometric Median]
\label{prp:geomMed}
Let $\dist = N(\zero, \SSigma)$ be a distribution with diagonal covariance matrix $\SSigma$ whose variance along the coordinate direction $\ee_1$ is zero, and equal to $1$ in all the other coordinate directions. Assume there is an $\eta$  fraction of noise at a distance $a = n$ along $\ee_1$. Let 
\begin{equation}
\label{eq:geomMed}
 t_0 = \arg \min_t (1-\eta) \E_{\xx \sim \dist}\pr{ \sqrt{t^2 + x_2^2 + ... + x_n^2} } + \eta (a-t).
 \end{equation}
Then, $t = \Omega(\eta \sqrt{n}).$
\end{proposition}
\proof 
We have that at the minimizer $t_0$, the derivative with respect to $t$ is zero. Therefore, we should have
$$ \E_{\xx \sim \dist} \frac{t_0}{\sqrt{t_0^2 + x_2^2 + ... + x_n^2 }} = \frac{\eta}{1-\eta} .$$
Consider $f(t) = \E_{\xx \sim \dist} \frac{t}{\sqrt{t^2 + x_2^2 + ... + x_n^2 }}.$  It is clear from Equation~\ref{eq:geomMed} that $t_0 >0.$ We claim that  if $t = \alpha \eta \sqrt{n}$ for a small enough constant $\alpha$, then $f(t) \leq \frac{\eta}{1-\eta}.$ Suppose $t_1 = \alpha \eta \sqrt{n}$. Since $\xx \sim \dist$, 
$\| \xx \|_2^2 \geq n/2$ with exponential probability. Therefore, 
\begin{align*}
f(t_1) &\leq  \E_{\xx \sim \dist} \frac{t_1}{\sqrt{t_1^2 + n/2 }} \\
& \leq \frac{t_1 \sqrt{2 \pi}}{\sqrt{t_1^2 + n/2 }} \leq {\alpha \eta  \sqrt{2 \pi}}.
\end{align*}
The claim, and hence the proof follows.

\qed

\subsection{Algorithms}
Our algorithms are based on outlier removal and SVD. To simplify the proofs, we use new samples for each step of the algorithm. The total sample complexity is given in the theorems.
\subsubsection{Outlier Removal}
For outlier removal, we use one of the following two simple routines. The first, which we call {\em OutlierDamping}, returns a vector of positive weights, one for each sample point.
\begin{framed}
\captionof{algocf}{\label{fig:OutlierDamping} $\outdamp(S)$ \hfill \hfill}
Input: $S \subset \rea^{n}$ with $|S| = m$  \\
Output: $S \subset \rea^{n}, \ww = (w_1,...,w_m) \in \rea^{m} $ 
\begin{enumerate}
\item \IF $n=1$: \\
\quad  \RETURN $(S,-1).$
\item Let $\aa$ be the coordinate-wise median of $S$. Let $s^2 = C \Tr(\SSigma).$ Estimate $\Tr(\SSigma)$ by estimating $1$d variance along $n$ orthogonal directions, see Section~\ref{sec:1dCovEst}.
\item Set $w_i = \exp \left( - \frac{ \| \xx_i - \aa \| _2^2}{s^2} \right)$ for every $\xx_i \in S.$
\item \RETURN $(S,\ww).$
\end{enumerate}
\end{framed}

The second procedure for outlier removal returns a subset of points. It will be convenient to view this as a $0/1$ weighting of the point set. We call this procedure {\em OutlierTruncation}.

\begin{framed}
\captionof{algocf}{\label{fig:OutlierTruncation} $\outrunc(S, \eta)$ \hfill \hfill}
Input: $S \subset \rea^{n}, \eta \in [0,1]$   \\
Output: $\widetilde{S} \subset S, \ww = \one \in \rea^{m} $ 
\begin{enumerate}
\item \IF $n=1$: \\
\quad Let $[a,b]$ be the smallest interval containing $(1-\eta-\eps)(1-\eta)$ fraction of the points, $\widetilde{S} \leftarrow S \cap [a,b]$. \RETURN $(\widetilde{S},1).$
\item Let $\aa$ be as in Lemma~\ref{lem:easyMean}.
\item Let $B(r,\aa)=$ ball of minimum radius $r$ centered at $\aa$ that contains $(1-\eta-\eps)(1-\eta)$ fraction of $S$.
\item $\widetilde{S} \leftarrow S \cap B(r,\aa).$  \RETURN $(\widetilde{S},\one).$
\end{enumerate}
\end{framed}

\subsubsection{Main Algorithm}
We are now ready to state the main algorithm for agnostic mean estimation. It uses one of the above outlier removal procedures and assumes that the output of the procedure is a weighting.

\begin{framed}
\captionof{algocf}{\label{fig:pca}  \pca($S$)  \hfill \hfill}
Input: $S \subset \mathbb{R}^n,$ and a routine $\outrem(\cdot)$.\\ 
Output: $\mmuhat \in \mathbb{R}^n.$ 
\begin{enumerate}
\item Let $ ( \widetilde{S}, \ww) = \outrem(S)$ . 
\item {\IF $n=1$:
\begin{enumerate}
\item{\IF $\ww = -1$, \RETURN $\med(\widetilde{S})$. //Gaussian case} \item{\ELSE \RETURN $\mean(\widetilde{S})$. //General case}
\end{enumerate} }
\item  Let $\SSigma_{\widetilde{S},\ww}$ be the weighted covariance matrix of $\widetilde{S}$ with weights $\ww$, and $V$ be the span of the top $n/2$ principal components of $\SSigma_{\widetilde{S},\ww}$, and $W$ be its complement.
\item Set $S_1 := \PP_V (S)$ where $\PP_V$ is the projection operation on to $V$.
\item Let $\mmuhat_V := \pca(S_1)$ and $\mmuhat_{W} := \text{mean} (\PP_{W} \widetilde{S}).$
\item Let $\mmuhat \in  \mathbb{R}^n$ be such that 
$\PP_V \mmuhat =  \mmuhat_V$ and 
$ \PP_{W} \mmuhat = \mmuhat_{W}.$
\item \RETURN $\mmuhat.$
\end{enumerate}
\end{framed}

\subsubsection{Estimation of the Covariance Matrix and Operator Norm}

For both the tasks in this section, we will assume that the mean of the distribution $\mmu = \zero$. We can do this without loss of generality by a standard trick mentioned described in Section~\ref{sec:mainCovEst}.  The algorithm for estimating the covariance matrix calls $\pca$ on $\xx \xx ^T.$  Analysis is given in Section~\ref{sec:mainCovEst}. 
 \begin{framed}
\captionof{algocf}{\label{fig:covEst} \covEst(S)  \hfill \hfill}
Input: $S \subset \mathbb{R}^n, \eta \in \rea$  \\
Output: $n \times n$ matrix $\SSigmahat$ 
\begin{enumerate}
\item Let $S^{(2)} = \{  \xx_i' \xx_i' | \, i = 1,...,m/2\}$  (see Equation~\ref{eq:sym})
\item Run the mean estimation algorithm on $S^{(2)}$, where elements of $S^{(2)}$ are viewed as vectors in $\rea^{n^2}.$ Let the output be $\SSigmahat$.
\item \RETURN $\SSigmahat.$
\end{enumerate}

\end{framed}

 The algorithm for estimating $\| \SSigma \|_2$  is based on iteratively truncating the samples along the direction of top variance. The analysis is given in Section~\ref{sec:main2norm}.

\begin{framed}
\captionof{algocf}{\label{fig:powit} \powit(S)  \hfill \hfill}
Input: $S \subset \mathbb{R}^n, \eta \in [0,1],\gamma \in \mathbb{R}$  \\
Output: $\sigma^2 \in \rea_{>0} .$ 
\begin{enumerate}
\item Let $\widetilde{S} = \safeoutrunc(S, \eta,\gamma)$. \label{alg:outlierstep}
\item Do the following $O(n \log^{2/\gamma} \frac{n}{\eta})$ times
\item Let $\SSigma_{\zero} (\widetilde{S}):= \frac{1}{|\widetilde{S}|} \sum_{i \in \widetilde{S}} \xx \xx^T$.\label{alg:iterstep}
\item Find $\vv$, the top eigenvector of $\SSigma_{\zero}(\widetilde{S})$, and its corresponding eigenvalue $\sigma^2$.
\item Estimate (up to $1 \pm c\eta$ factor, see Section~\ref{sec:1dCovEst}) the variance of $\dist$ along $\vv$ and denote it by $\sigmahat_{\vv}^2$.\label{alg:estStep}
\item \IF $\sigma^2 \leq (1 + c_3 \eta \log^{2/\gamma} \frac{n}{\eta}) \sigmahat_{\vv}^2$
\\ \quad \RETURN $\sigma^2$.
\item Remove all points $\xx\in \widetilde{S}$ such that $|\xx^T\vv| > \frac{c_2 \sigmahat_{\vv} \log^{1/\gamma} \frac{n}{\eta}}{2}$. \label{alg:truncstep}

\item Go to Step (\ref{alg:iterstep}). 
\end{enumerate}
\end{framed}

\begin{framed}
	\captionof{algocf}{\label{fig:SafeOutlierTruncation} $\safeoutrunc(S, \eta, \gamma)$ \hfill \hfill}
	Input: $S \subset \rea^{n}, \eta \in [0,1],\gamma \in \rea$   \\
	Output: $\widetilde{S} \subset S$ 
	\begin{enumerate}
		\item Let $t = \sum_{i=1}^n \widehat{\sigma}^2_{e_i}$ be the sum of estimated variances of $\dist$ in $n$ orthogonal directions.
		\item Let $B(c \sqrt{t}\log^{1/\gamma} \frac{n}{\eta},\zero)$ be the ball of radius $c \sqrt{t}\log^{1/\gamma} \frac{n}{\eta}$ centered at $\zero$.
		\item $\widetilde{S} \leftarrow S \cap B(c\sqrt{t}\log^{1/\gamma} \frac{n}{\eta},\zero).$  \RETURN $\widetilde{S}.$
	\end{enumerate}
\end{framed}

\section{Mean Estimation: Theorem~\ref{thm:mainGauss} and Theorem~\ref{thm:mainGeneral}}
In this section, we will first prove Theorem~\ref{thm:mainGauss}, which is for Gaussian distributions, and  Theorem~\ref{thm:mainGeneral}, which is for distributions with bounded fourth moments. All our algorithms will be translationally invariant. We will assume  w.l.o.g  that the mean of the distribution $\dist$ is $\mmu=0$. So we will be proving bounds on $\| \mmuhat \|_2.$ Algorithm \ref{fig:pca} has $\log n$ levels, we will assume that at each level it uses $O(\frac{n \log n}{\eps^2})$ samples resulting in a total of $m = O(\frac{n \log^2 n}{\eps^2}).$

At various points in the analysis, to bound the sample complexity we will have to show that the estimates computed from samples are close to their expectations. We will use the following two results. 
Firstly, as an immediate corollary of matrix Bernstien for rectangular matrices (see Theorem $1.6$ in \cite{Tropp2012}), we get the following concentration result for the sample mean and sample covariance.
\begin{lemma}
\label{lem:covConcentration}
Consider a distribution in $\R^n$ with covariance matrix $\SSigma$ and supported in some Euclidean ball whose radius we denote is $\sqrt{R \| \SSigma \|},$ for some $R \in \rea.$ Let $\epsilon \in (0,1).$  Then the following holds with probability at least $1 - 1/\pol(n)$: $\text{If } N \geq \frac{R \log n}{\epsilon^2} \text{ then }$
$$ \| \widehat{\mmu} - \mmu \| \leq \epsilon \sqrt{\| \SSigma \|}$$
 and
$$ \| \widehat{\SSigma} - \SSigma \| \leq \epsilon \| \SSigma \|.$$
Here $\widehat{\mu}$ and  $\widehat{\Sigma}$ are sample mean and  sample covariance matrix.
\end{lemma}

Secondly, the functions we estimate will be integrals of low-degree polynomials (degree $d$ at most $4$) restricted to intervals and/or balls. These functions viewed as binary concepts have small VC-dimension, $O(n^d)$ where $n$ is the dimension of space and $d$ is the degree of the polynomial. We use this to bound the error of estimating integrals via samples, and we can make the error smaller than any inverse polynomial using a $\pol(n)$ size sample. 
\begin{proposition}
\label{prop:vcIntegration}
Let $F$ be a class of real-valued functions from $\R^n$ to $[-R, R]$. Let $C_F$ be the corresponding class of binary concepts, i.e., for each $f \in F$, we consider the concepts $h_t(x) = 1$ if $f(x)\ge t$ and zero otherwise. Suppose the VC-dimension of $C_F$ is $d$. Then, for any $f \in F$, and any distribution $\dist$ over $\R^n$, an iid sample $S$ of size $|S| \ge \frac{8}{\eps^2}\left(d \log (1/\eps) + \log(1/\delta)\right)$, with probability at least $1-\delta$ satisfies
\[
\left|\E_{x \sim \dist}(f(x)) -   \frac{1}{|S|}\sum_{x \in S} f(x) \right|  \le 2\eps R.
\]
\end{proposition}
\proof 
By the VC theorem, for any concept in $C_F$, the bound on the size of the sample ensures that with probability at least $1-\delta$ and any $t$, 
\[
\left| \Pr(f(x) \ge t) - \frac{|\{x \in S \, : \, f(x)\ge t\}|}{|S|}\right| \le \eps.
\]
Noting that 
$\E_{x \sim \dist}(f(x)) = \int_{-R}^R \Pr(f(x) \ge t) \, dt$, we get the claimed bound.

\qed


 Let $s^2 := \frac{1}{\epsilon_1}  \Tr (\SSigma)$ and $\epsilon_2 := \frac{ \| \aa \| _2^2}{\eta^ 2 s^2}$. We can estimate $\Tr(\SSigma)$ by estimating ($1$ dimensional) variances along $n$ orthogonal directions, see Section~\ref{sec:1dCovEst}. Note that we can arrange $0< \epsilon_1,\epsilon_2 <1 $ to be small enough constants. We weight every point $\xx$ by $w_{\xx} = \exp ( - \frac{ \| \xx - \aa \| ^2}{s^2}).$
Let $\dist = N(0,\SSigma)$ be a Gaussian distribution and  $S = \{ \xx_1,...,\xx_m \}, \xx_i \sim \dist_{\eta}$ be the sample we get. Let $S = S_G \cup S_N$ be the Gaussian and the noise points repectively, with $|S_N| = \eta m$. For a set $T \subset \rea^n$, let
\begin{align*}
\mmu_{T,\ww}  := \frac{1}{m}\sum_{i \in T} w_{\xx_i} \xx_i \, \, \text{ and }\, \, \SSigma_{T, \ww} := \frac{1}{|T|} \sum_{i \in T} w_i (\xx_i - \mmu_{T,\ww} )(\xx_i - \mmu_{T,\ww} )^T 
\end{align*}
We use the above notation for $T = S_G$ and $T = S_N$. By an abuse of notation, when $T=G$, we mean the population version of the above quantities:
\begin{align*}
\mmu_{G,\ww}  := \E_{\xx} w_{\xx} \xx \, \, \text{ and }\, \, \SSigma_{G, \ww} := \E_{\xx} w_{\xx} (\xx - \mmu_{G,\ww} )(\xx_i - \mmu_{G,\ww} )^T. 
\end{align*}

 Note that 
$$\mmu_{S, \ww} = (1-\eta)\mmu_{S_G,\ww} + \eta \mmu_{S_N,\ww}.$$

We  consider the  matrix 
$\SSigma_{S, \ww} $ 
\begin{align*}
 \SSigma_{S, \ww} &=
 \frac{1}{m} \sum_i w_{\xx_i} (\xx_i - \mmu_{S,\ww} ) (\xx_i - \mmu_{S,\ww} )^T \\
& = (1- \eta)\SSigma_{S_G, \ww}  + \eta \SSigma_{S_N, \ww} + \eta (1- \eta) (\mmu_{S_N,\ww}  -  \mmu_{S_G,\ww})(\mmu_{S_N,\ww}  -  \mmu_{S_G,\ww})^T.
\end{align*}

\subsection{\Proofof{Theorem \ref{thm:mainGauss}}}
\label{sec:mainGauss}
%
Let us assume $\eta < 1/2.1$. We then have 
\begin{lemma}
\label{lem:1dGauss}
Let $\dist = N(0, \sigma^2)$ be a one dimensional Gaussian distribution. If $m = O\pr{ \frac{\log n} {\eps^2} }$, and we are given $x_1,...,x_m \sim \dist_{\eta}$, then the median $x_{\text{med}} = \med_i \{x_i \}$ satisfies $|x_{\text{med}}| = O((\eta + \eps) \sigma)$ with high probability.
\end{lemma}
\proof 
Let $S_G \subset S$ be made up of samples in $S$ that come from the Gaussian, also let $c = \Phi^{-1}(1/2 + \eta + \eps).$ Let us bound the probability that the median $x_{\text{med}} \geq c.$ We first note that if  
$x_{\text{med}} \geq c,$ then $\Pr \pr{ x >c | x \in_u S_G} \geq \eps. $ By Hoeffding's inequality, we can bound this by $1 - \pol(n)$ if 
$|S_G| = O\pr{\frac{\log n}{\eps^2}}.$
\qed

We will next consider the multidimensional case. The proof follows by a series of lemmas. We state the lemmas first, conclude the proof of Theorem~\ref{thm:mainGauss} and then prove the lemmas.  First, we observe that by applying Lemma \ref{lem:1dGauss} in $n$ orthogonal directions and union bound, we get

\begin{lemma}
Suppose $\vv_1,...,\vv_n \in \rea^n$ are a set of  orthonormal vectors.  Suppose $m_i = \med_j \{ \vv_i^t \xx_j\},$ and $\aa = \sum_i m_i \vv_i.$ Then if $m = O\pr{\frac{\log n}{ \eps^2}},$ there exists a constant $C$ independent of the choice of $\vv_i$'s such that with probability $1- \pol(n)$ ,
$$ \| \aa  \| ^2_2 \leq C \eta^2 \Tr (\SSigma). $$
\end{lemma}

 By a simple calculation,  $ \max_{\xx} \| \xx \|^2 e^{- \| \xx - \aa \|^2/s^2} \leq O(s^2).$ This immediately gives the following bound on the trace. 
\begin{lemma}
\label{lem:traceBound}
Suppose $\AA := \eta \SSigma_{S_N, \ww} + \eta (1- \eta) (\mmu_{S_N,\ww}  -  \mmu_{G,\ww})(\mmu_{S_N,\ww}  -  \mmu_{S_G,\ww})^T.$
Then there exists a constant $C$  such that, $$\Tr(\AA) \leq C \eta s^2.$$
\end{lemma}
 We will show later
\begin{theorem}
\label{thm:covBoundGauss}
  $$ \left( \frac{e ^{-\eta^2 \epsilon_2}}{1 + \epsilon_1} - \eta^2 \epsilon_2 e^{2 \epsilon_1} \right)  \SSigma \preceq  \SSigma_{G, \ww}  \preceq e^{\epsilon_1} \SSigma. $$ 
  \end{theorem}
As will be clear from the proof of Theorem~\ref{thm:covBoundGauss}, when $\SSigma = \sigma^2 \II$ is a multiple of identity, then 
$ \SSigma_{G, \ww} $ will also be a multiple of $\II$.  
By Lemma~\ref{lem:covConcentration}, if we take $m = O(\frac{ n \log n}{\epsilon^2})$ samples, we will have 
$$ (1 - \epsilon) \SSigma_{G, \ww} \preceq \SSigma_{S_G, \ww} \preceq (1 + \epsilon) \SSigma_{G, \ww}.$$
Suppose, we have
$$\alpha \SSigma \preceq \SSigma_{S_G, \ww} \preceq \beta \SSigma$$ 
in the Lowener ordering, for some $\alpha, \beta > 0$.
By an argument similar to the one sketched in Section \ref{sec:mainIdeas}, we can prove

\begin{lemma}
\label{lem:eigBoundGaus}
We will use the notation as defined above. Let $W$ be the bottom $n/2$ principal components of the covariance matrix $ \SSigma_{S, \ww} $. We have
$$\|\eta P_W  \ddelta_{\mmu} \|^2  \leq 2\eta \left( (\beta + C \eta) \| \SSigma \| _2  - \alpha  \| \SSigma \| _{\min} \right),$$
where  $\| \SSigma \|_{\min}$ denotes the least eigenvalue of $\SSigma$ and $\ddelta_{\mmu} := \mmu_{S_N,\ww}  -  \mmu_{S_G,\ww}.$
\end{lemma}

By an inductive application of Lemma~\ref{lem:eigBoundGaus}, we get the following theorem giving a bound on $\| \mmuhat \|.$
\begin{theorem}
\label{thm:main}
On input $S$ and the routine $\outdamp(\cdot)$, $\pca$ outputs $\mmuhat$ satisfying 
$$ \|  \mmuhat  \| ^2 \leq O\pr{  (\beta \eta +  \eta^2 + \eps^2) \| \SSigma \| _2  - \alpha \eta  \| \SSigma \| _{\min} } (1 + \log n).$$ 
\end{theorem}

Theorem~\ref{thm:covBoundGauss} combined with Theorem~\ref{thm:main} proves Theorem~\ref{thm:mainGauss}. We get a better dependence on $\eta$ when $\SSigma = \sigma^2 \II$ because we can take $\alpha = \beta$ in this case. This would lead to the cancellation of the leading term in the bound in Theorem~\ref{thm:main} as $\| \SSigma \| _2 = \| \SSigma \| _{\min}.$

\qed

\Proofof{Lemma \ref{lem:eigBoundGaus}}
Recall that $\SSigma$ denotes the covariance matrix of the Gaussian part. We have
\begin{align*}
\SSigma_{S, \ww} & = (1- \eta)\SSigma_{S_G, \ww}  + \eta \SSigma_{S_N, \ww} + \eta (1- \eta)  \ddelta_{\mmu} \ddelta_{\mmu}^T \\
& = (1 - \eta)\SSigma_{S_G, \ww} + \AA,
\end{align*}
where $\AA = \eta \SSigma_{S_N, \ww} + \eta (1- \eta)  \ddelta_{\mmu} \ddelta_{\mmu}^T.$
Therefore, we have
$$(1 - \eta)\alpha{\SSigma}  + \AA \preceq \SSigma_{S, \ww}  \preceq (1 - \eta)\beta{\SSigma}  + \AA.$$

For a symmetric matrix $\BB$, let $\lambda_k(\BB)$ denote the $k$'th largest eigenvalue. By Weyl's inequality, we have  
$$ \lambda_k( (1 - \eta)\SSigma_{G, \ww} + \AA ) \leq \lambda_k(\AA) + (1-\eta)\beta  \| \SSigma \| _2.$$
 Therefore,  
 $$ \lambda_{n/2} \left( \SSigma_{S, \ww}  \right) \leq \lambda_{n/2} \left( \AA \right) + (1-\eta)\beta \| \SSigma \| _2.$$
 By Lemma~\ref{lem:traceBound} we have
 \begin{align*}
 \lambda_{n/2} \left( \AA \right) &\leq \frac{\Tr(\AA)}{n/2} \\
 & \leq 2C^2 \eta  \| \SSigma \| _2 \\
 \implies \lambda_{n/2}(\SSigma_{S, \ww} ) &\leq (1-\eta)\beta \| \SSigma \| _2  + 2C^2 \eta  \| \SSigma \| _2 \\
& \leq (\beta + 2 C^2 \eta) \| \SSigma \| _2.
 \end{align*}

Recall that $W$ is the space spanned by the bottom $n/2$ eigenvectors of $\SSigma_{S, \ww} $, and $\PP_W$ is the matrix corresponding to the projection operator on to $W$. We therefore have
$$ \PP_W ^T \SSigma_{S, \ww}  \PP_W \preceq (\beta + 2 C^2 \eta) \| \SSigma \| _2 \II.$$
We therefore have
$$ \alpha \PP_W^T \SSigma \PP_W +  \eta \PP_W^T  \SSigma_{S_N, \ww} \PP_W+ (\eta - \eta^2) (\PP_W  \ddelta_{\mmu})( \PP_W   \ddelta_{\mmu})^T \preceq (\beta + 2 C^2 \eta) \| \SSigma \| _2  \II.$$
Multiplying by the vector $\frac{\PP_W  \ddelta_{\mmu}}{ \| \PP_W   \ddelta_{\mmu} \| }$ and its transpose on either side, we get 
$$(\eta-\eta^2) \| \PP_W   \ddelta_{\mmu} \| ^2 \leq (\beta + 2 C^2 \eta) \| \SSigma \| _2  - \alpha  \| \SSigma \| _{\min}.$$
Assuming $\eta \leq 1/2$, we therefore have 
$$ \| \eta \PP_W  \ddelta_{\mmu} \| ^2 \leq 2\eta \left( (\beta + 2 C^2 \eta) \| \SSigma \| _2  - \alpha  \| \SSigma \| _{\min} \right).$$
\qed

\Proofof{Theorem \ref{thm:main}}  By Equation~\ref{eq:meanBound} and Lemma \ref{lem:covConcentration}, since we take $O\pr{ \frac{n \log n}{\eps^2} }$ samples we have 
\begin{align*}
\|  \mmu_{S_G,\ww} \| _2^2 &\leq \pr{ \eta^2 \epsilon_2 e^{2 \epsilon_1}  + \eps^2} \| \SSigma \| _2 \\
& = O\pr{\eta^2  + \eps^2}   \| \SSigma \| _2.
\end{align*}
So it is enough to prove 
$  \|  \mmuhat - \mmu_{S_G,\ww} \| ^2 \leq O\pr{  (\beta \eta +  \eta^2 + \eps^2) \| \SSigma \| _2  - \alpha \eta  \| \SSigma \| _{\min} } (1 + \log n)$
The proof is by induction. If $n=1$, then the conclusion follows from the guarantees of the one dimensional median. Now, assume that it holds for all $n \leq k$ for some $k \geq 1$. Let $n = k+1.$ We have by Lemma \ref{lem:eigBoundGaus} 
\begin{align*}
  \| \eta \PP_W \left( \mmu_{S_N,\ww}  -  \mmu_{S_G,\ww} \right)  \| ^2 &\leq    O \left( (\beta \eta +  \eta^2) \| \SSigma \| _2  - \alpha \eta  \| \SSigma \| _{\min} \right) \\
 \implies  \|  \PP_W \mmu_{S,\ww} - \PP_W \mmu_{S_G,\ww} \| _2^2 &\leq  O\left( (\beta \eta +  \eta^2) \| \SSigma \| _2  - \alpha \eta  \| \SSigma \| _{\min} \right).
\end{align*}

  By induction hypothesis, since $\text{dim}(V) = n/2$, we have 
 $$ \|  {\mmuhat_V} - \PP_V  \mmu_{S_G,\ww} \| ^2 \leq O\pr{  (\beta \eta +  \eta^2 + \eps^2) \| \SSigma \| _2  - \alpha \eta  \| \SSigma \| _{\min} } (1 + \log n/2) .$$ 
 Therefore, adding the two, we get 
 $$ \|   \mmuhat - \mmu_{S_G,\ww} \| ^2 \leq  O\pr{  (\beta \eta +  \eta^2 + \eps^2) \| \SSigma \| _2  - \alpha \eta  \| \SSigma \| _{\min} } (1 + \log n/2)  .$$
 \qed
 


 \Proofof{Theorem \ref{thm:covBoundGauss}}

%
We will first consider the second moment 
$$B := \E_{\xx} \exp \pr{ - \frac{ \| \xx - \aa \| ^2}{s^2} } \xx \xx^T.$$
We have
\begin{align*}
B & = \frac{1}{\sqrt{ (2 \pi)^n |\SSigma| }} \int \exp \left( - \frac{ \| \xx - \aa \| ^2}{s^2} \right) \exp \pr{-\xx^T \SSigma^{-1} \xx } \xx \xx^T dx \\
& =  \frac{1}{\sqrt{ (2 \pi)^n |\SSigma| }} \int \exp \left( - \frac{ \| \xx - \aa \| ^2}{s^2} -\xx^T \SSigma^{-1} \xx \right) \xx \xx^T dx \\
& = \frac{1}{\sqrt{ (2 \pi)^n |\SSigma| }} \exp \left( -\frac{ \| \aa \| ^2}{s^2}  + \frac{1}{s^4} \aa^T \left( \SSigma^{-1} + \frac{1}{s^2}\II \right)^{-1} \aa \right)\\& \int \exp \left(- (\xx - \bb )^T \left( \SSigma^{-1} + \frac{1}{s^2}\II \right) (\xx - \bb  )  \right) \xx \xx^T dx,
\end{align*}

where $\bb = \frac{1}{s^2} \left( \SSigma^{-1} + \frac{1}{s^2}\II \right)^{-1} \aa.$ Therefore, we have 
$$ B = \exp \left( -\frac{ \| \aa \| ^2}{s^2}  + \frac{1}{s^4} \aa^T \left( \SSigma^{-1} + \frac{1}{s^2}\II \right)^{-1} \aa \right) \frac{1}{| \SSigma | \left| \SSigma^{-1} + \frac{1}{s^2}\II \right| } \left( \SSigma^{-1} + \frac{1}{s^2}\II \right)^{-1}.$$

Now we will look at the scalar term $ | \SSigma | \left| \SSigma^{-1} + \frac{1}{s^2}\II \right|.$ Let $\lambda_i$ be the eigenvalues of $\SSigma.$
\begin{align*}
 | \SSigma | \left| \SSigma^{-1} + \frac{1}{s^2}\II \right| &= \Pi_i  \frac{\left| \frac{1}{\lambda_i} + \frac{1}{s^2} \right|}{\frac{1}{\lambda_i}} \\
 & = \Pi_i \left( 1 + \frac{\lambda_i}{s^2} \right). \\
 \end{align*}

We then have 
$$ 1 + \epsilon_1 \leq | \SSigma | \left| \SSigma^{-1} + \frac{1}{s^2}\II \right| \leq e^{\epsilon_1}.$$

We next bound $\exp \left( -\frac{ \| \aa \| ^2}{s^2}  + \frac{1}{s^4} \aa^T \left( \SSigma^{-1} + \frac{1}{s^2}\II \right)^{-1} \aa \right).$ 
We  have 
$$ \frac{1}{s^4} \aa^T \left( \SSigma^{-1} + \frac{1}{s^2}\II \right)^{-1} \aa  \leq  \frac{1}{s^2} \aa^T \aa .$$
Therefore
$$ \exp (-\eta^2 \epsilon_2) \leq \exp \left( -\frac{ \| \aa \| ^2}{s^2}  + \frac{1}{s^4} \aa^T \left( \SSigma^{-1} + \frac{1}{s^2}\II \right)^{-1} \aa \right) \leq 1.$$

Therefore, 
$$ e ^{-\eta^2 \epsilon_2} \left( \SSigma^{-1} + \frac{1}{s^2}\II \right)^{-1}  \preceq  B \preceq e^{\epsilon_1} \left( \SSigma^{-1} + \frac{1}{s^2}\II \right)^{-1}  .$$

\begin{lemma}
\label{lem:matBound}
We have the following
$$\frac{1}{1 + \epsilon_1}  \SSigma \preceq \left( \SSigma^{-1} + \frac{1}{s^2}\II \right)^{-1}  \preceq \SSigma$$
\end{lemma}
\proof
Note that if $\frac{1}{\lambda_1},...,\frac{1}{\lambda_n}$ and $\vv_1,...,\vv_n$ are the eigenvalues and the corresponding eigenvectors of $\SSigma^{-1}$, then $\frac{1}{\lambda_1} + \frac{1}{s^2},...,\frac{1}{\lambda_n} + \frac{1}{s^2}$ and  $\vv_1,...,\vv_n$ are the eigenvalues and the corresponding eigenvectors of  $\SSigma^{-1} + \frac{1}{s^2}\II.$ Since,
$$ \frac{\lambda_i}{1 + \epsilon_1} \leq \frac{1}{\frac{1}{\lambda_i} + \frac{1}{s^2}} \leq \lambda_i$$
the lemma follows.
\qed

From Lemma~\ref{lem:matBound}, we have
\begin{equation}
\label{eq:covBound}
\frac{e ^{-\eta^2 \epsilon_2}}{1 + \epsilon_1}  \SSigma \preceq B \preceq e^{\epsilon_1} \SSigma.
\end{equation}
Next we will bound $$ \mmu_{G,\ww} = \E_{\xx} w_{\xx} \xx .$$

\begin{align*}
 \mmu_{G,\ww} & = \frac{1}{\sqrt{ (2 \pi)^n |\SSigma| }} \int \exp \left( - \frac{ \| \xx - \aa \| ^2}{s^2} \right) \exp(-\xx^T \SSigma^{-1} \xx) \xx dx \\
& =  \frac{1}{\sqrt{ (2 \pi)^n |\SSigma| }} \int \exp \left( - \frac{ \| \xx - \aa \| ^2}{s^2} -\xx^T \SSigma^{-1} \xx \right) \xx dx \\
& = \frac{1}{\sqrt{ (2 \pi)^n |\SSigma| }} \exp \left( -\frac{ \| \aa \| ^2}{s^2}  + \frac{1}{s^4} \aa^T \left( \SSigma^{-1} + \frac{1}{s^2}\II \right)^{-1} \aa \right)\\& \int \exp \left(- (\xx - \bb )^T \left( \SSigma^{-1} + \frac{1}{s^2}\II \right) (\xx - \bb  )  \right) \xx dx \\
&= \exp \left( -\frac{ \| \aa \| ^2}{s^2}  + \frac{1}{s^4} \aa^T \left( \SSigma^{-1} + \frac{1}{s^2}\II \right)^{-1} \aa \right) \frac{1}{| \SSigma | \left| \SSigma^{-1} + \frac{1}{s^2}\II \right| }   \bb ,
\end{align*}
where $\bb = \frac{1}{s^2} \left( \SSigma^{-1} + \frac{1}{s^2}\II \right)^{-1} \aa.$ Recall that $\epsilon_1 = \frac{\sum_i \lambda_i}{s^2}$. We can, as before, bound the product of the two scalars by $e^{\epsilon_1}.$ Therefore, we have 

$$ \|  \mmu_{G,\ww} \| _2 \leq e^{\epsilon_1} \left|   \frac{1}{s^2} \left( \SSigma^{-1} + \frac{1}{s^2}\II \right)^{-1} \aa \right|.$$

Therefore, we have
\begin{align*}
 \| \mmu_{G, \ww} \| _2^2 & = e^{2\epsilon_1} \frac{1}{s^4} \aa^T \left( \SSigma^{-1} + \frac{1}{s^2}\II \right)^{-1/2} \left( \SSigma^{-1} + \frac{1}{s^2}\II \right)^{-1} \left( \SSigma^{-1} + \frac{1}{s^2}\II \right)^{-1/2}\aa \\
& \leq e^{2\epsilon_1} \frac{1}{s^2} \aa^T \left( \SSigma^{-1} + \frac{1}{s^2}\II \right)^{-1} \aa \\
& \leq e^{2\epsilon_1} \frac{\aa^T \aa}{s^2} \left \|  \SSigma^{-1} + \frac{1}{s^2}\II \right \| ^{-1}_2  \\
& \leq  \eta^2 \epsilon_2 e^{2\epsilon_1} \left \|  \SSigma^{-1} + \frac{1}{s^2}\II \right \| ^{-1}_2 \\
& = \eta^2 \epsilon_2 e^{2\epsilon_1} \frac{1}{1/ \| \SSigma \| _2 + 1/s^2}\\
&\leq \eta^2 \epsilon_2 e^{2\epsilon_1}  \| \SSigma \| _2.
\end{align*}

Also, similarly
\begin{equation*}
 \|  \mmu_{G,\ww} \| _{ \SSigma^{-1} + \frac{1}{s^2}\II }^2 \leq  \eta^2 \epsilon_2 e^{2 \epsilon_1}.
\end{equation*}

This implies 

\begin{equation*}
\mmu_{G,\ww} \mmu_{G,\ww}^T \preceq  \eta^2 \epsilon_2 e^{2 \epsilon_1} \left( \SSigma^{-1} + \frac{1}{s^2}\II \right)^{-1} .
\end{equation*}

From Lemma~\ref{lem:matBound}, we have 
\begin{equation}
\label{eq:meanBound}
 \mmu_{G,\ww} \mmu_{G,\ww}^T \preceq  \eta^2 \epsilon_2 e^{2 \epsilon_1} \SSigma .
\end{equation}

Combining Equation \eqref{eq:meanBound} and Equation \ref{eq:covBound}, we get Theorem \ref{thm:covBoundGauss}.

\qed

\subsection{Improving the dependence on $\eta$}
\label{sec:improvedEta}
Now we will show how we can obtain the second part of Theorem~\ref{thm:mainGauss} to get a better dependence on $\eta$ by using $\SSigmahat$ from Theorem~\ref{thm:mainCovEst}. Let $\dist = N(\mmu, \SSigma)$ be a Gaussian with covariance $\SSigma,$ and $\eta \leq  c/\log n$ for a small enough constant $c >0.$ We first use Theorem~\ref{thm:mainCovEst} (with $\eps = \eta$) to estimate $ \sigma^2 = \| \SSigma \|_2$. We get a $\sigmahat^2$ satisfying
\begin{equation}
\label{eq:2norm}
\pr{1 - O(\sqrt{\eta \log n}) } \sigma^2 \leq \sigmahat^2 \leq \pr{1 + O(\sqrt{\eta \log n}) } \sigma^2.
\end{equation}
Let $S = \{ \xx_1,...,\xx_m \}$ be the given sample, and let $\yy_i \sim N(0, \sigmahat^2 \II), i=1,...,m$ be i.i.d. samples. Define $\xx_i' = \xx_i + \yy_i.$ The key thing to note is that  if $\xx \sim N(\mmu, \SSigma) $ and $\yy \sim N(0, \sigmahat^2 \II)$, then $\xx + \yy \sim N(\mmu, \SSigma + \sigmahat^2 \II)$. Let $\dist' = N(\mmu, \SSigma + \sigmahat^2 \II)$. Note that the mean $\mmu'$ of $\dist'$ is same as that of  $\dist$, and the covariance $\SSigma' =   \SSigma + \sigmahat^2 \II$ has 
\begin{equation}
\label{eq:eigbounds}
\lambda_{\max}\pr{\SSigma'} \leq  \pr{2 + O(\sqrt{\eta \log n}) } \sigma^2 \text{ and  } \lambda_{\min} \pr{\SSigma'}  \geq \pr{1 - O(\sqrt{\eta \log n}) } \sigma^2.
\end{equation}

We can view $\xx_i' \sim \dist'_{\eta},$ and we assume $\eta \log n \leq c.$ By Theorem~\ref{thm:mainCovEst} and Equation~\ref{eq:2norm}, we can compute a $\SSigmahat'$ such that 
$$ \left \| \SSigmahat' - \SSigma' \right \|_F \leq O \left( \sqrt{\eta \log n } \right ) \sigma^2  .$$
Let $\alpha = O \left( \sqrt{\eta \log n  }    \right )   .$ Therefore,

\begin{align*}
  & \SSigmahat' - \alpha \sigma^2 \II \preceq \SSigma' \preceq \SSigmahat' + \alpha \sigma^2 \II \\
 & \implies \II - \alpha \sigma^2 \SSigmahat'^{-1} \preceq \SSigmahat'^{-1/2} \SSigma' \SSigmahat'^{-1/2} \preceq \II + \alpha \sigma^2 \SSigmahat'^{-1} \\
 & \implies \pr {1-O \left( \sqrt{\eta \log n  } \right ) } \II  \preceq \SSigmahat'^{-1/2} \SSigma' \SSigmahat'^{-1/2} \preceq \pr{1+ O \left( \sqrt{\eta \log n  } \right) }  \II   
\end{align*}
by Equation~\ref{eq:eigbounds}.
Now, if we let $\xx_i'' = \SSigmahat'^{-1/2} \xx_i'$ and $\dist'' = N(\mmu'', \SSigma'') = N \pr{ \SSigmahat'^{-1/2} \mmu,  \SSigmahat'^{-1/2} \SSigma \SSigmahat'^{-1/2}}, $ then we can think of $\xx_i'' \sim \dist''_{\eta}.$ If we now use Theorem~\ref{thm:main} with $\beta = \pr{1+ O \left( \sqrt{\eta \log n  } \right) }$ and $\alpha = \pr{1- O \left( \sqrt{\eta \log n  } \right)}$  on the samples $S'' = \{ \xx_i'' \}$, we get a $\mmuhat''$ such that 
$$ \| \mmuhat'' - \mmu'' \|^2 = O(\eta^{3/2} \log^{3/2} n).$$ 

This implies that $\mmuhat =\SSigmahat'^{1/2} \mmuhat''$ satisfies
\begin{align*}
 \| \mmuhat - \mmu \|^2 &= O(\| \SSigmahat' \| \eta^{3/2} \log^{3/2} n)\\
 &= O(\| \SSigma \|_2 \eta^{3/2} \log^{3/2} n).
\end{align*}

\begin{remark}
We can use this technique to give a polynomial time algorithm to compute $\mmuhat$ with a guarantee $ \| \mmuhat - \mmu \|^2 = O \pr{ \| \SSigma \|_2 \eta^{2 - \epsilon} \log^{2 - \epsilon} n }$ for any fixed $\epsilon >0.$ This would require estimating higher order moments by the mean estimation algorithm and then using the above trick to improve the $\eta$ dependence for each of them in sequence. We don't give a proof of this in this paper.
\end{remark}

\subsection{Distributions with Bounded Fourth Moments}
\label{sec:mainGeneral}
In this section, we will prove some some useful properties that distributions with bounded fourth moments satisfy. We will assume that $\xx \sim \dist$ for a distribution with mean $\mmu$ that has bounded fourth moments, i.e., for every unit vector $\vv$
\begin{equation}
\label{eq:4thmom}
 \E ((\xx-\mmu)^T\vv)^4 \leq \const \left( \E ((\xx-\mmu)^T\vv)^2 \right)^2 ,
 \end{equation}
for some $\const.$

\begin{lemma}[Mean shift]
\label{lem:meanShift}
Let $X$ be a random variable with $\E (X-\E X)^2 = \sigma^2$ and  $$\E (X-\E X)^4 \leq \const \pr{ \E (X-EX)^2 }^2,$$
for some $\const$. Let $\epsilon \leq 0.5$ and $A$ be any event with probability $\Pr(A) = 1 - \epsilon.$ Then 
$$\left | \E (X | A) - \E(X) \right | \leq \sqrt[4]{8\const \epsilon^3} \sigma.$$
\end{lemma}
\proof Let $a = \E (X | A)$. Then
\begin{align*}
\E X &= (1-\epsilon)a + \epsilon \E(X|A^c)\\
\iff \E(X|A^c) &= \frac{\E X - (1-\epsilon)a}{\epsilon} =  \frac{1-\epsilon}{\epsilon}(\E X - a) + \E X
\end{align*}
The fourth moment of such an $X$ is minimum when its support is  just the two-point set $\{ a, \frac{1-\epsilon}{\epsilon}(\E X - a) + \E X \}$. Therefore,

\begin{align*}
& (1-\epsilon) (a-\E X)^4 +  \epsilon \left(\frac{1-\epsilon}{\epsilon}(\E X-a) \right)^4 \leq \const \sigma^4 \\
\implies &\abs{a-\E X} \leq \sqrt[4]{\frac{\const \epsilon^3}{(1-\epsilon)(3\epsilon^2 - 3 \epsilon + 1)}} \sigma \leq \sqrt[4]{8\const \epsilon^3} \sigma,
\end{align*}
when $\epsilon \leq 0.5.$
\qed

\begin{lemma}
\label{lem:2ndMomShift}
Let $X$ be a random variable with $\E X = \mu$ and $\E((X-\mu)^2) = \sigma^2$ and let
$$\E (X-\mu)^4 \leq \const \sigma^2,$$
for some $\const$. Then, for every event $A$ that occurs with probability at least $1- \epsilon$, we have 
\begin{equation}
\label{eq:2ndmom1}
\E \left( (X-\mu)^2 \one_A \right) \geq \pr{1-\sqrt{\const\epsilon}} \sigma^2,
\end{equation}
where $\one_A$ is the indicator function of the event $A.$ As an immediate corollary, for $\epsilon \le 0.5$ we get the following bound on the conditional probability
$$ - \sqrt{\const \epsilon}\sigma^2 \le \E \left( (X-\mu)^2 | A \right) - \sigma^2 \leq 2\epsilon \sigma^2 .$$

 \end{lemma}
\proof  Let $d \Omega$ be the probability measure. We can write $\E (X-\mu)^4 \leq \const (\E (X-\mu)^2)^2$ in the following way
\begin{align*}
\int_{A} (X-\mu)^4 d \Omega + \int_{A^c} (X-\mu)^4 d \Omega  &\leq \const \left( \int_{A} (X-\mu)^2 d \Omega + \int_{A^c} (X-\mu)^2 d \Omega \right)^2
\end{align*}
Using $\E (Y- \E Y)^4 \geq (\E (Y-\E Y)^2)^2$  for any random variable $Y,$ and $\Pr(A^c) = \epsilon$ we have    
$$ \frac{1}{ \epsilon } \left ( \int_{A^c} (X-\mu)^2 d \Omega\right)^2 \leq \int_{A^c} (X-\mu)^4 d \Omega$$
We therefore have
\begin{align*}
\frac{\left ( \int_{A^c} (X-\mu)^2 d \Omega\right)^2}{\epsilon} &\leq  \const \left( \int_{A} (X-\mu)^2 d \Omega + \int_{A^c} (X-\mu)^2 d \Omega \right)^2\\
\iff  \int_{A^c} (X-\mu)^2 d \Omega &\leq  \sqrt{\const\epsilon} \left( \int_{A} (X-\mu)^2 d \Omega + \int_{A^c} (X-\mu)^2 d \Omega \right)\\
\iff \pr{1-\sqrt{\const\epsilon}} \left(\int_{A^c} (X-\mu)^2 d \Omega\right) &+ \pr{1-\sqrt{\const\epsilon}} \left(\int_{A} (X-\mu)^2 d \Omega\right) \leq  \int_{A} (X-\mu)^2 d \Omega \\
\iff  \pr{1-\sqrt{\const\epsilon}} \E (X-\mu)^2 &\leq  \int_{A} (X-\mu)^2 d \Omega
\end{align*}

This proves the inequality~\eqref{eq:2ndmom1}. Now, 
\begin{align*}
\E \left( (X-\mu)^2 | A \right) &= \frac{1}{\mu(A)}\int_{A} (X-\mu)^2 d \Omega \\
& \geq \pr{1-\sqrt{\const\epsilon}} \sigma^2.
\end{align*}
Also, 
\begin{align*}
\E \left( (X-\mu)^2 | A \right) &= \frac{1}{\mu(A)}\int_{A} (X-\mu)^2 d \Omega \\
& \leq \frac{1}{1 - \epsilon} \sigma^2.
\end{align*}
Therefore, for $\epsilon \le 0.5$ we get that 
$$ - \sqrt{\const \epsilon}\sigma^2 \le \E \left( (X-\mu)^2 | A \right) - \sigma^2 \leq 2\epsilon \sigma^2 .$$

\qed

As an immediate corollary of Lemma~\ref{lem:meanShift} and Lemma~\ref{lem:2ndMomShift}, we get for a random variable $\xx$ having bounded fourth moments
\begin{corollary}
\label{cor:varBound} Let $A$ be an event that happens with probability $1 - \eta$.  Then, 
$$\pr{1 - O(\sqrt{\const\eta})} \SSigma \preceq \SSigma|_{A} \preceq (1 + 2\eta) \SSigma ,$$
where $\SSigma|_A$ is the conditional covariance matrix
$\SSigma|_A := \E (\xx \xx^T|A) - (\E(\xx|A))(\E(\xx|A))^T.$
\end{corollary}
\proof
Let $\vv$ be any unit vector. Let $y$ be the random variable that is $\vv^T \xx$ for $\xx\sim \dist$. Let $\mu = \E(y)$, $\mu_A = \E (y|A)$, and $d = \mu_A - \mu$. Then
\begin{align*}
\E ((y - \mu_A)^2|A) = \E ((y - \mu - d)^2|A) &= \E((y-\mu)^2|A) - 2 d \E(y - \mu | A) + d^2\\
&= \E((y-\mu)^2|A) - d^2
\end{align*}
By Lemma $\ref{lem:meanShift}$ and Lemma $\ref{lem:2ndMomShift}$, 
\begin{align*}
\E ((y - \mu_A)^2|A) - \E ( (y-\mu)^2) &\le 2\eta \E((y-\mu)^2)\\
\E ((y - \mu_A)^2|A) - \E ((y-\mu)^2) &\ge - \sqrt{\const\eta} \E ((y-\mu)^2) - d^2\\
&\ge -\pr{\sqrt{\const\eta}  + \sqrt{8\const\eta^3}} \E ((y-\mu)^2)
\end{align*}

\qed

Finally, by a standard argument as in the proof of Chebyshev's inequality, we have
 
\begin{lemma}[Concentration]
\label{lem:concentration}
For every unit vector $\vv$, we have
$$\Pr \pr{|\xx^T \vv - \E \xx^T \vv| \geq t \sigma_{\vv} } \leq \frac{\const}{t^4},$$
where $\sigma_{\vv}$ is the standard deviation of $\xx$ along the direction $\vv$, $\sigma_{\vv}^2 :=  \E |\xx^T \vv|^2 - |\E \xx^T \vv|^2.$
\end{lemma}

\subsection{\Proofof{Theorem \ref{thm:mainGeneral}}}
\label{sec:4thmomout}
\subsubsection{One Dimensional Distribution}
\label{sec:4thmom1d}
First we will consider the case when $X$ is a random variable with mean $\mu$ and variance $\sigma^2$ satisfying 
$$ \E ((X-\mu)^4) \leq \const \sigma^4 .$$
In this case, median need not be a good estimator. Instead, we will consider the interval of minimum length that contains $(1 - \eta - \eps)(1- \eta)$ fraction of the sample points. Let $S$ be the given sample, and let $\widetilde{S}$ be the points lying in this interval. Let $\widehat{\mu} = \text{mean}(\widetilde{S})$ be our estimator. We will show below that $|\widehat{\mu} - \mu| \leq O \pr{\const^{1/4} (\eta + \eps)^{3/4} \sigma }.$

By the concentration inequality stated in Lemma~\ref{lem:concentration}, we get that for the distribution, the length $r_{1-\frac{\eta + \eps}{2}}$ of the interval around $\mu$ consisting of probability mass $1 - \frac{\eta + \eps}{2} $ is bounded by 
\begin{align*}
r_{1-\frac{\eta + \eps}{2}} \leq \frac{\const^{1/4}}{\pr{\frac{\eta + \eps}{2}}^{1/4}}\sigma.
\end{align*}
We  will refer to this interval by $I_{1-\frac{\eta + \eps}{2}}$. We note that by VC theorem if  $ \abs{S_{\dist}} = \Omega \pr{ \frac{\log n + \log 1/\eps}{\eps^2} }$, then with probability $1 - 1/\pol(n)$ for every interval $I \subset \rea$, 
$$\left| \Pr \pr{ x \in I \, | \, x \sim \dist } - \Pr \pr{ x \in I \, | \, x \, \eps_{u} \, S_{\dist} } \right | \leq \eps/2.$$
\anup{Include VC proof?}
The length of the smallest interval that contains $(1 - \eta - \eps)(1-\eta)$ fraction of $S$ is at most the length of the smallest interval that contains $1-\eta - \eps$ fraction of $S_{\dist}$. This latter quantity is bounded by $r_{1-\eta}$, since the interval $I_{1-\frac{\eta + \eps}{2}}$ contains with  probability $1 - 1/\pol(n)$ a $(1 - \eta - \eps)$ fraction of $S_{\dist}$. 

This implies that when we look at the minimum interval containing $1 - \eta - \eps$ fraction of the non-noise points, the extreme points of the interval can be at most at a distance $ r _{1-\frac{\eta + \eps}{2}} $ from $\mu$. Thus, the distance of all noise points will be within $O\left(\frac{\const^{1/4}}{(\eta + \eps)^{1/4}}\sigma \right)$. Furthermore, the interval of minimum length with $(1-\eta - \eps)(1-\eta)$ fraction of $S$ will contain at least $1 - 3 \eta - \eps$ fraction of $S_{\dist}$. Therefore, by Lemma~\ref{lem:meanShift} the mean of $\widetilde{S}$ will be within $\eta \cdot r_{1-\eta}+ O\pr{ \sqrt[4]{\const (\eta + \eps)^3} \sigma} = O \pr{\const^{1/4} (\eta + \epsilon)^{3/4} \sigma }$ from the true mean.

\subsubsection{Multi-dimensional Case}

We will now consider the multidimensional case. Let $\dist$ be a distribution on $\rea^n$ and $\xx \sim \dist$ is a random variable that satisfies for every direction $\vv$
$$ \E (((\xx-\mmu)^T\vv)^4) \leq \const \left( \E (((\xx-\mmu)^T\vv)^2) \right)^2 ,$$
for some $\const.$

For any direction $\vv$, let $\mu_v = \mmu^T v$. From the previous section, we know that we can find a $\widehat{\mu}_{\vv}$ such that 
$$ |\widehat{\mu}_{\vv} - \mu_v| \leq O(\const^{1/4} (\eta + \eps)^{3/4} \sigma_{\vv}).$$
Therefore, by picking $n$ orthogonal directions $\vv_1,...,\vv_n$, we get 
\begin{lemma}
\label{lem:easyMean}
Given $O\pr{ \frac{n \log n}{\eps^2} }$ samples, we can find a vector  $\aa \in \rea^n$ such that $ \| \aa - \mmu \| _2 = O(\const^{1/4} (\eta + \eps)^{3/4} \sqrt{ \Tr(\SSigma)}).$
\end{lemma}

We will now bound the radius of the ball in the outlier removal step (Algorithm \ref{fig:OutlierTruncation}). We claim the radius of the ball is $O\pr{\frac{\const^{1/4}}{(\eta+\eps)^{1/4}}\sqrt{n||\SSigma||_2}}$. Suppose we have some $\xx \sim \dist$. Let $\zz = \xx - \mmu$. Using the $n$ orthogonal directions as picked above, let $z_i = \zz^T \vv_i$ and let $Z^2 = \sum z_i^2 = \norm{\zz}_2^2$. Consider the following:
\begin{align}
\Pr\pr{Z^2 \ge \frac{\const^{1/2} n||\SSigma||_2}{(\eta + \eps)^{1/2}} }  = \Pr\pr{Z^4 \ge \frac{\const n^2||\SSigma||^2_2}{\eta + \eps} }  \le \frac{(\eta + \eps)\E(Z^4)}{\const n^2||\SSigma||^2_2} \label{eq:outlier}
\end{align}
It suffices to bound the right-hand side of $(\ref{eq:outlier})$ by $O(\eta + \eps)$, in which case the ball will contain $1-\eta - \eps$ fraction of the probability mass of $\dist$. We have
\begin{align}
\E(Z^4)= \E\pr{\sum_i z_i^2 \sum_j z_j^2} \le n^2\max_i \E \pr{z_i^4} \le C_4 n^2 \| \SSigma \|_2^2
\end{align}
due to the fourth moment condition and the fact that $\E((\zz^T \vv_i)^2)\le \| \SSigma \|_2$. Therefore, a ball of radius at most $O\pr{\frac{\const^{1/4}}{(\eta + \eps)^{1/4}}\sqrt{n||\SSigma||_2}}$ contains $1 - \eta - \eps$ fraction of the points. Since $ \| \aa-\mmu \| _2 = O\pr{\const^{1/4}(\eta + \eps)^{3/4}\sqrt{\Tr(\SSigma)}},$ we get that the radius of the ball computed in the outlier removal step is $O\left( \frac{\const^{1/4}}{(\eta + \eps)^{1/4}}  \sqrt{n \|\SSigma \|_2} \right).$ \anup{Add a VC statement?}We have proved

\begin{lemma}
\label{lem:noiseRad}
After the outlier removal step, every remaining point $\xx$ satisfies
$$  \| \xx-\mmu \| _2 \leq O\left( \frac{\const^{1/4}}{(\eta+\eps)^{1/4}}  \sqrt{n \|\SSigma \|_2} \right).$$
\end{lemma}

 Consider the covariance matrix $\SSigma_{\widetilde{S}}$ of $\widetilde{S}$ (recall that $\widetilde{S}$ is the sample after outlier removal). Let  $\widetilde{S}_{\dist} \subset \widetilde{S}$ be the set of points in $\widetilde{S}$ that were sampled from the distribution $\dist$ and $\widetilde{S}_N \subset \widetilde{S}$ be the points sampled by the adversary. Let $\mmu_{\widetilde{S}} := \text{mean}(\widetilde{S})$,  $\mmu_{\widetilde{S}_N}:=\text{mean}(\widetilde{S}_N)  $ and $\mmu_{\widetilde{S}_\dist}:=\text{mean}(\widetilde{S}_{\dist}) .$ Note that 
 $$\mmu_{\widetilde{S}} = \widetilde{\eta} \mmu_{\widetilde{S}_N}  + (1 - \widetilde{\eta}) \mmu_{\widetilde{S}_\dist},$$
 where $\widetilde{\eta} = \frac{|\widetilde{S}_N|}{|\widetilde{S}|}$ is the fraction of noise points after the outlier truncation step. Note that $\widetilde{\eta} \leq \frac{\eta}{1-2 \eta - \eps} = O(\eta).$ We will therefore pretend that the fraction of noise points is still $\eta$ after the outlier truncation step.
We again assume that the mean of the distribution $\dist$ is $\mmu = 0.$ By Lemma~\ref{lem:meanShift} applied with $X = \xx^T\frac{\mmu_{\widetilde{\dist}}}{\|  \mmu_{\widetilde{\dist}} \|}$ for $\xx \sim \dist$ and where $A$ is the event that $\xx$ is not removed by outlier removal, we have that 
\begin{equation}
\label{eq:meanShift}
\|  \mmu_{\widetilde{\dist}} \|_2 = O(\const^{1/4} (\eta + \eps)^{3/4} \sqrt{\| \SSigma \|_2}).
\end{equation}
 
 \anup{rephrase this}Suppose, after the outlier removal step, we had the guarantee that the covariance matrix of the remaining points from the distribution $\dist$, say $\SSigma_{{\widetilde{\dist}}},$ is between 
$$\alpha (1-\eta) \SSigma \preceq \SSigma_{{\widetilde{\dist}}} \preceq \beta (1-\eta) \SSigma$$ 
in the Lowener ordering.  Corollary~\ref{cor:varBound} gives 
$\alpha = 1-O(\sqrt{\const (\eta + \eps)})$ and $\beta = 1 + O(\eta + \eps)$.
Also, by Lemma~\ref{lem:covConcentration} and Lemma~\ref{lem:noiseRad} we have that if $| S_{\widetilde{\dist}} | = \Omega \pr{\frac{n \log n}{\eps^2}}$, then 
$$ (1 - \frac{\const^{1/4}\eps}{(\eta + \eps)^{1/4}}) \SSigma_{{\widetilde{\dist}}}   \preceq  \SSigma_{S_{\widetilde{\dist}}}   \preceq (1 + \frac{\const^{1/4}\eps}{(\eta + \eps)^{1/4}}) \SSigma_{{\widetilde{\dist}}}$$
We will use the notation as defined above.
\begin{lemma}
\label{lem:eigBound}
 Let $W$ be the bottom $n/2$ principal components of the covariance matrix $\SSigma_{S}$. For some constant $C$, we have
$$ \| \eta \PP_W  \ddelta_{\mmu}  \| ^2  \leq O \pr{  (\beta \eta + \const^{1/2} \eta^{3/2}) \| \SSigma \| _2  - \alpha \eta  \| \SSigma \| _{\min} },$$
where $\ddelta_{\mmu} =  \mmu_{\widetilde{S}_N}  -  \mmu_{\widetilde{S}_{\dist}}.$
\end{lemma}

By an inductive application of the above lemma, we can prove
\begin{theorem}
\label{thm:main2}
On input $(S,n)$, $\pca$ outputs $\mmuhat$ satisfying 
$$ \|   \mmuhat \| ^2 \leq O \pr{  (\beta \eta + \const^{1/2} (\eta + \eps)^{3/2}) \| \SSigma \| _2  - \alpha \eta  \| \SSigma \| _{\min} } (1 + \log n).$$ 
\end{theorem}

 Theorem~\ref{thm:main2} with Corollary~\ref{cor:varBound} proves Theorem~\ref{thm:mainGeneral}. 

\Proofof{Lemma \ref{lem:eigBound}}
Recall that $\SSigma$ denotes the covariance matrix of the points from $\dist$. We have
\begin{align*}
\SSigma_{\widetilde{S}} &=   (1 - \eta)\SSigma_{\widetilde{S}_{\dist}} + \eta \SSigma_{\widetilde{S}_N}  + (\eta - \eta^2)\ddelta_{\mmu} \ddelta_{\mmu}^T \\
& = (1 - \eta)\SSigma_{\widetilde{S}_{\dist}} + \AA,
\end{align*}
where $\AA := \eta \SSigma_{{\widetilde{S}_N}} + (\eta - \eta^2)\ddelta_{\mmu} \ddelta_{\mmu}^T  .$
Therefore, we have
$$(1 - \eta)\alpha{\SSigma}  + \AA \preceq \SSigma_{\widetilde{S}}  \preceq (1 - \eta)\beta{\SSigma}  + \AA.$$
By Lemma~\ref{lem:noiseRad} each $\xx_i$ satisfies 
$ \|  \xx_i \|  = O\left( \frac{\const^{1/4}}{(\eta+\eps)^{1/4}}  \sqrt{n\|\SSigma\|_2} \right),$ 
so we have
\begin{equation}
\label{eq:tracebound}
\Tr(\AA) = O\pr{\frac{\eta \sqrt{\const}\|\SSigma \|_2 n}{\sqrt{\eta+\eps}}} \le O\pr{\sqrt{\const \eta}  \| \SSigma \| _2  n}. 
\end{equation}

For a symmetric matrix $B$, let $\lambda_k(B)$ denote the $k$'th largest eigenvalue. By Weyl's inequality, we have that 
$$ \lambda_k( (1 - \eta) \SSigma_{\widetilde{S}} +\AA ) \leq \lambda_k(\AA) + (1-\eta)\beta  \| \SSigma \| _2.$$
 Therefore,  
 $$ \lambda_{n/2} \left( \SSigma_{\widetilde{S}} \right) \leq \lambda_{n/2} \left( \AA \right) + (1-\eta)\beta \| \SSigma \| _2.$$
 By Equation~\eqref{eq:tracebound}, there exists a constant $\widetilde{C}$ such that
 \begin{align*}
 \lambda_{n/2} \left( \AA \right) &\leq \frac{\Tr(\AA)}{n/2} \\
 & \leq \widetilde{C} \sqrt{\const \eta}  \| \SSigma \| _2,
 \end{align*}
we have, 
\begin{align*}
\lambda_{n/2}(\SSigma_{\widetilde{S}}) &\leq (1-\eta)\beta \| \SSigma \| _2  + \widetilde{C} \sqrt{\const \eta}  \| \SSigma \| _2
\end{align*}
Recall that $W$ is the space spanned by the bottom $n/2$ eigenvectors of $\SSigma_{\widetilde{S}}$, and $\PP_W$ is the matrix corresponding to the projection operator on to $W$. We therefore have
\begin{align*}
\PP_W ^T \SSigma_{\widetilde{S}} \PP_W &\preceq ((1-\eta)\beta + \widetilde{C} \sqrt{\const \eta}) \| \SSigma \| _2 \II\\
(1-\eta)\alpha \PP_W^T \SSigma \PP_W +  \eta \PP_W^T \SSigma_{\widetilde{S}_N} \PP_W +& (\eta - \eta^2) (\PP_W  \ddelta_{\mmu})( \PP_W  \ddelta_{\mmu})^T \preceq ((1-\eta)\beta + \widetilde{C} \sqrt{\const \eta}) \| \SSigma \| _2  \II
\end{align*}

Multiplying by the vector $\frac{\PP_W  \ddelta_{\mmu}}{ \| \PP_W  \ddelta_{\mmu} \| }$ and its transpose on either side, we get 
$$\eta \| \PP_W \ddelta_{\mmu} \| ^2 \leq (\beta + C \sqrt{\const \eta}) \| \SSigma \| _2  - \alpha  \| \SSigma \| _{\min}.$$
where $C = \frac{\widetilde{C}}{1-\eta}$. We therefore have 
$$ \| \eta \PP_W \ddelta_{\mmu} \| ^2 \leq \eta \left( (\beta + C \sqrt{\const \eta}) \| \SSigma \| _2  - \alpha  \| \SSigma \| _{\min} \right).$$
\qed

\Proofof{Theorem \ref{thm:main2}} By Equation~\ref{eq:meanShift}, it is enough to bound $ \|   \mmuhat - \mmu_{\widetilde{S}_{\dist}}\| ^2.$ The proof is by induction on the dimension. If $n=1$, then the conclusion follows from the guarantees for the one dimensional case proven in Section~\ref{sec:4thmom1d}. Now, assume that it holds for all $n \leq k$ for some $k \geq 1$. Let $n = k+1.$ We have by Lemma \ref{lem:eigBound} 
\begin{align*}
\|  \PP_W \left( \mmuhat - \mmu_{\widetilde{S}_{\dist}} \right ) \|^2 & = \|  \eta \PP_W \ddelta_{\mmu}  \| ^2 \\
& \leq O \pr{  (\beta \eta + \const^{1/2} \eta ^{3/2}) \| \SSigma \| _2  - \alpha \eta  \| \SSigma \| _{\min} }
\end{align*}

Recall that we defined $V$ to be the span of the top $n/2$ principal components of $\SSigma_{\widetilde{S}}$. By induction hypothesis, since $\text{dim}(V) = n/2$, we have 
 $$ \|  \mmuhat_{V} - \PP_V  \mmu_{\widetilde{S}_{\dist}} \| ^2 \leq O \pr{  (\beta \eta + \const^{1/2} (\eta + \eps)^{3/2}) \| \SSigma \| _2  - \alpha \eta  \| \SSigma \| _{\min} }
  (1+\log n/2) .$$ 
 
 Therefore, adding the two, we get 
 $$ \|  \mmuhat -  \mmu_{\widetilde{S}_{\dist}}\| ^2 \leq  O \pr{  (\beta \eta + \const^{1/2} (\eta + \eps)^{3/2}) \| \SSigma \| _2  - \alpha \eta  \| \SSigma \| _{\min} }
  (1+\log n) .$$
 \qed

\section{Covariance Estimation}

\subsection{One Dimensional Case}
\label{sec:1dCovEst}
\begin{observation}[$1$d Covariance Estimate]
\begin{enumerate}
\item Let $\dist$ be a distribution with mean $\mmu$ and covariance $\sigma^2.$ If $\dist = N(\mu, \sigma^2)$, then there is an algorithm that takes as input $m = \O \pr{ \frac{\log n}{\eps^2}}$ samples $\xx_1,...\xx_m \sim \dist_{\eta}$ and computes in polynomial time  $\widehat{\sigma}^2$ such that $\left |\widehat{\sigma}^2 - \sigma^2 \right | = O(\eta + \eps )\sigma^2.$ 
\item If $\xx \sim \dist$ has bounded fourth moments with constant $\const$, and $(x - \mu)^2$ has bounded fourth moments with constant $C_{4,2}$.  Then there is an algorithm that takes as input $\eta$ and $m =O \pr{\frac{\log n + \log 1/\eps}{\eps^2}}$ samples $\xx_1,...\xx_m \sim \dist_{\eta}$ and computes in polynomial time  $\widehat{\sigma}^2$ such that $\left |\widehat{\sigma}^2 - \sigma^2 \right | = O \pr{ C_{4,2}^{1/4} (\eta + \eps)^{3/4} \const^{1/2} \sigma }.$
\end{enumerate}
 \end{observation}
 
 \proof 
 
When the distribution $\dist$ is supported on $\rea,$ we can estimate the variance in the following way. We will consider the case $\dist = N(\mu,\sigma)$ and $\dist$ just having bounded eighth moments separately. Suppose $\dist = N(\mu, \sigma)$, and we are given $m = \pol(n)$ samples $S = \{ x_1,...,x_m\}, x_i \sim \dist_{\eta}.$ There are several ways to estimate $\sigma$, we describe here one of them. First we  compute the median, and let $x_{\text{med}} = \med_i \{ x_i \}.$ 
Let $\Phi(x)$ be the c.d.f. of $N(0,1)$. Note that $c_{1} = \Phi(1) \sim 85.1$. Let $C_{\sigma}$ be the $c_{1}$'th quantile of $S.$ Then our estimate for the standard deviation is $\widehat{\sigma} = C_{\sigma} - \widehat{\mu}.$ By Lemma~\ref{lem:1dGauss}, we have $\left | \widehat{\mu} - \mu \right | = O(\eta \sigma).$  For a similar reason, $C_{\sigma} = \sigma \pm O(\eta \sigma).$ Therefore,  
$\widehat{\sigma^2} = \sigma^2 \pm O(C_{8}^{1/4}\eta \sigma^2).$

When $\dist$ is a distribution that has bounded eighth moments, the result follows from the 1d mean estimation in Section~\ref{sec:4thmomout} applied $(x - \mu)^2$. Note that $\E (x - \mu)^2 = \sigma^2$ and 
\begin{align*}
\E \pr{ (x - \mu)^2  -\sigma^2 }^2 &= \E (x - \mu)^4 - \sigma^4 \\
& \leq \const \sigma^4.
\end{align*}
From Section~\ref{sec:4thmomout}, we therefore have that if $m = O \pr{\frac{\log n + \log 1/\eps}{\eps^2}}$, there is a $\pol(n)$ algorithm with 
$|\widehat{\sigma}^2 - \sigma^2| \leq O \pr{ C_{4,2}^{1/4} (\eta + \eps)^{3/4} \const^{1/2} \sigma }.$

\qed

\subsection{Multi-Dimensional Case: {Theorem \ref{thm:mainCovEst}}}
\label{sec:mainCovEst}
In this section we will prove that $\covEst$ (Algorithm~\ref{fig:covEst}) gives Theorem~\ref{thm:mainCovEst}. 
Throughout this section, we will assume that  $\dist$ is a distribution with mean $\mmu$ and covariance $\SSigma$ and has bounded fourth moments with parameter $C_4$. We use the following symmetrization trick to assume that $\dist$ has mean $\zero$. Given samples $S = \{ \xx_1,...,\xx_m \}$, let 
\begin{equation}
\label{eq:sym}
\xx'_i = \frac{\xx_i - \xx_{i+m/2}}{\sqrt{2}} \text{ for } i \in \{1,...,m/2\}.
\end{equation}
Since $\eta$ fraction of the original samples were corrupted on average, only $2\eta$ fraction of the new samples will be corrupted on average. Moreover, if $\xx, \yy \sim \dist$ are independent random variables, then we can show that the distribution of $\xx' =  (\xx - \yy)/\sqrt{2}$  has bounded fourth moments with parameter $\leq C_4 + 3/2$.  We will denote by $\dist'$ the distribution of $\xx'$. $\covEst$ is just the mean estimation algorithm on  $S^{(2)} = \{ \xx' \xx'^T  | \xx \in S \}$, we can appeal to Theorem~\ref{thm:mainGeneral}. Furthermore, let $\dist'$ be an affine transformation of a $4$-wise independent distribution.


  Note that 
$$\E_{\xx \sim \dist'} \xx \xx ^T = \SSigma .$$
By Theorem~\ref{thm:mainGeneral}, we have 
$$\| \SSigmahat - \SSigma  \|_F = O  \pr{\eta^{1/2} + C_{4,2}^{1/4} (\eta + \eps)^{3/4} }\| \SSigmac \|^{1/2}_2 \log^{1/2} n, $$
where $\SSigmac$ is covariance matrix of $\xx \xx ^T, \xx \sim \dist'.$

By Proposition~\ref{prp:mainCovariance} , we have 
$$\| \SSigmahat - \SSigma  \|_F = O  \pr{\eta^{1/2} + C_{4,2}^{1/4} (\eta + \eps)^{3/4} }  \const^{1/2} \|\SSigma \|_2 \log^{1/2} n,$$
which proves Theorem~\ref{thm:mainCovEst}.

\qed

  We will now derive a bound for $ \| \SSigmac \|_2$ when the distribution has bounded fourth moments and is $4$-wise independent. In particular, we will prove 
\begin{proposition}
\label{prp:mainCovariance}
If $\SSigmac$ is the covariance matrix of $\xx \xx ^T, \xx \sim \dist'$, it holds that 
 $$ \| \SSigmac \| _2 \leq  O \left(  \const  \| \SSigma \|^2 _2 \right ).$$
\end{proposition}

\Proofof{Proposition~\ref{prp:mainCovariance}}
Note that $\E(\YY)=\SSigma$.  
\begin{align*}
 \E(((\YY-\E(\YY))\cdot V)^2) &= \E \left(\sum_{ij}(\YY_{ij}-\SSigma_{ij})V_{ij}\right)^2 \\
&= \sum_{ijkl}\E((\YY_{ij}-\SSigma_{ij})(\YY_{kl}-\SSigma_{kl}))V_{ij}V_{kl}\\
&= \sum_{ijkl}\E(\YY_{ij}\YY_{kl}-\SSigma_{ij}\SSigma_{kl})V_{ij}V_{kl}\\
&=\sum_{ijkl}\E(x_ix_jx_kx_l -\SSigma_{ij}\SSigma_{kl})V_{ij}V_{kl}.
\end{align*}
Next we note that 
\[
\E(x_ix_jx_kx_l) -\SSigma_{ij}\SSigma_{kl}= \begin{cases} \E(x_i^4)-\SSigma_{ii}^2 \mbox{ if } i=j=k=l\\
\E(x_i^2x_j^2)  \mbox{ if } i=k, j=l \mbox{ or } i=l, j=k\\
0 \mbox{ otherwise}.
\end{cases}
\]
Therefore, 
\begin{align*}
\max_{V: \|V\|_F = 1} \E(((\YY-\E(\YY))\cdot V)^2) &= \max_{V: \|V\|_F=1} \sum_{i}(\E(x_i^4)-\SSigma_{ii}^2)V_{ii}^2 + 2\sum_{i < j}\SSigma_{ii}\SSigma_{jj} V_{ij}^2\\
&= \max_{V: \|V\|_F = 1} \sum_i (\E(x_i^4)-2\SSigma_{ii}^2)V_{ii}^2 + \sum_{i,j} \SSigma_{ii}\SSigma_{jj}V_{ij}^2\\
&\le \max_i \E(x_i^4)-2\SSigma_{ii}^2   + \max_{i} \SSigma_{ii}^2.\\
&\le  O\pr{\const} \|\SSigma\|_2^2.
\end{align*}

\qed

\section{Estimating $\| \SSigma \|_2 $: {Theorem \ref{thm:main2norm}}}
\label{sec:main2norm}

As in Section \ref{sec:mainCovEst}, we assume that the true distribution has mean $\mmu = \zero$. 


In this section, we will prove $\powit$ (Algorithm~\ref{fig:powit}) gives Theorem~\ref{thm:main2norm}. Let $S = S_\dist \cup S_N$ be the given sample, where $S_\dist$ consists of points from some distribution $\dist$ with mean $\mmu$ and covariance $\SSigma$ and $S_N$ consists of points picked by the adversary. Let $\SSigma_{S_\dist}$ be the sample covariance of $S_\dist$. We assume that $\dist$ has 1D concentration, i.e., there exists a constant $\gamma$ such that for every unit vector $\vv$
$$\Pr \left( \left |(\xx - \mmu)^T \vv \right| > t \sqrt{\vv^T\SSigma\vv} \right ) \leq e^{-t^{\gamma}}.$$

\subsection{Correctness}

Let $\widetilde{S}$ be the remaining sample at the end of the algorithm and let $\widetilde{S}_{\dist}$ be points in $\widetilde{S}$ sampled from $\dist$.

\begin{definition}
	Given a set of points $S \subset \rea^n$ and a vector $\aa \in \rea^n$, we let
	$$\SSigma_{\aa}(S) := \frac{1}{|S|} \sum_{\xx \in S} (\xx - \aa) (\xx - \aa)^T .$$
\end{definition}

First, we will argue that the covariance of the true distribution is well-approximated by $\SSigma_{\mmu}(\widetilde{S}_\dist)$.
\begin{lemma}\label{lem:wellapprox}
With probability $1-1/\pol(n)$,
$$ \| \SSigma- \SSigma_{\zero}(\widetilde{S}_\dist) \| \leq \eta \| \SSigma \| $$
\end{lemma}

\proof

%
%
 First, note that the $t$ computed in $\safeoutrunc$ is at most $O(\Tr(\SSigma))$ because by an analogous argument as in Section $\ref{sec:1dCovEst}$, we have $\widehat{\sigma}_{\vv}^2 \le (1+O(\eta))\sigma_{\vv}^2$ (namely that the estimated variance $\widehat{\sigma}_{\vv}$ in a direction $\vv$ is close to the true variance $\sigma_{\vv}$ in that direction). Then the ball in $\safeoutrunc$ has radius $R = c_1 \sqrt{\Tr(\SSigma)} \log^{1/\gamma} \frac{n}{\eta} $ for some constant $c_1$. We have that in any direction $\vv$, the probability that $\xx\sim \dist$ deviates from the mean by more than $c_1\sigma_{\vv} \log^{1/\gamma} \frac{n}{\eta}$ is $1/\pol(\frac{n}{\eta})$. Then if we take $n$ orthogonal directions, the probability that any given point is more than distance $R$  from $\mmu$ is still $1/\pol(\frac{n}{\eta})$. Thus, step (\ref{alg:outlierstep}) of the algorithm will remove only $1/\pol(\frac{n}{\eta})$ fraction of the points sampled from $\dist$.

In every direction $\vv$, the probability mass of points from $\dist$ outside an interval of size $ c_2 \sigma_{\vv} \log^{1/\gamma} \frac{n}{\eta}$ around the mean is at most $1/\pol(\frac{n}{\eta})$, where $\sigma_{\vv}$ is the variance in the direction $\vv$. Let $C_i$ be the region between the two hyperplanes used for truncation in iteration $i$. Therefore, if the number of iterations is $O(n \log^{2/\eta} \frac{n}{\eta})$, we will have that $\Pr \pr{\xx \in \cap_i C_i \left | x \sim \dist \right. } = 1 - 1/\pol(\frac{n}{\eta}).$ 

Note that $1$d concentration implies that the distribution has bounded $2k$'th moment for all finite $k.$ By  Lemma~\ref{lem:2ndMomShift}, we have that the covariance matrix 
$\SSigma_{\zero} \pr{\dist \cap_i C_i}$ of $\dist \cap_i C_i$ 
is close to that of $\SSigma$: 
\begin{equation}
\label{eq:2normtrunc}
(1 - 1/\pol(\frac{n}{\eta})) \SSigma \preceq  \SSigma_{\zero} \pr{\dist \cap_i C_i} \preceq (1 + 1/\pol(\frac{n}{\eta})) \SSigma.
\end{equation}
Finally, to relate $ \SSigma_{\zero} \pr{\dist \cap_i C_i}$ to $ \SSigma_{\zero} \pr{\widetilde{S}_{\dist}}$, we use Proposition~\ref{prop:vcIntegration}. The concept class we use is all degree two polynomials restricted to convex polytopes with at most $O(n)$ facets, defined by the hyperplanes used for truncation at each iteration of the algorithm. The VC dimension of this concept class is $O(n^2 \log n)$. Therefore,  by Proposition~\ref{prop:vcIntegration} applied with $R = c_1 \sqrt{\Tr(\SSigma)} \log^{1/\gamma} \frac{n}{\eta} \leq c_1 \| \SSigma \| ^{1/2} n^{1/2} \log^{1/\gamma} \frac{n}{\eta}$, we get that if we take 
$m = O \pr{ \frac{n^3 (\log^{1/\gamma} \frac{n}{\eta})^2 \log n}{\eta^2} } $ 
then 
\begin{equation}
\label{eq:vc2norm}
 \|  \SSigma_{\zero} \pr{\widetilde{S}_{\dist}} - \SSigma_{\zero} \pr{\dist \cap_i C_i} \| \leq \eta/2 \| \SSigma \|.
\end{equation}
Combining equations \ref{eq:2normtrunc} and \ref{eq:vc2norm} we get the desired result.
\qed

\begin{theorem} 
When the algorithm terminates, we have:
	$$  (1- \eta) \|\SSigma\|_2 \leq \|\SSigma_{\zero}(\widetilde{S})\|_2 \leq (1 + O( \eta \log^{2/\gamma} \frac{n}{\eta}))\|\SSigma\|_2.$$
	
\end{theorem}

\proof
First, note that since only an $\eta$ fraction of $\widetilde{S}$ is noise, we have
\begin{align}
\SSigma_{\zero}(\widetilde{S}) & \succeq (1 - \eta) \SSigma_{\zero}(\widetilde{S}_\dist) \label{eq:sigmanorm}
\end{align}

Therefore, we have that $\|\SSigma_{\zero}(\widetilde{S})\|_2 \geq (1- \eta) \|\SSigma_{\zero}(\widetilde{S}_\dist)\|_2.$ Lemma \ref{lem:wellapprox} gives the desired lower bound. For the upper bound, let $\vv$ be the top eigenvector of $\SSigma_{\zero}(\widetilde{S}).$ When the algorithm terminates, we have 
\begin{align*}
\|\SSigma_{\zero}(\widetilde{S})\|_2  &= \vv^T \SSigma_{\zero}(\widetilde{S})  \vv \\
&\leq (1 + O(\eta \log^{2/\gamma} \frac{n}{\eta}))\vv^T \SSigma \vv \\
& \leq (1 + O(\eta \log^{2/\gamma} \frac{n}{\eta}))\|\SSigma\|_2 .
\end{align*}
where the second line follows because of the termination condition and because we can estimate the variance of $\dist$ in any direction to within a $(1 \pm c\eta)$ factor.
\qed




\subsection{Termination}
In this section, we will show that with high probability, Algorithm~\ref{fig:powit} terminates in a polynomial number of steps provided that $\eta \le \frac{1}{C}$ for some constant $C$ that depends only on the estimation in Step (\ref{alg:estStep}).

Every time the algorithm goes through another iteration, it must remove a certain number of noise points. Suppose in step (\ref{alg:truncstep}), we remove $r$ noise points. The noise configuration of maximum variance puts $r$ amount of noise at the outlier removal distance $d_1 = c_1\sqrt{ \Tr(\SSigma)} \log^{1/\gamma} \frac{n}{\eta}$, and $\eta - r$ amount of noise at the truncation threshold distance $d_2 = \frac{c_2 \sigmahat_{\vv} \log^{1/\gamma} \frac{n}{\eta}}{2}$. We can then write an upper bound on $\sigma^2$.

\begin{align*}
\sigma^2 = \|(1-\eta)\SSigma_{\zero}(\widetilde{S}_\dist) + \eta\SSigma_{\zero}(\widetilde{S}_N)\|_2^2 \le \sigma_{\vv}^2 + rd_1^2 + (\eta - r)d_2^2
\end{align*}

This implies
\begin{align*}
r &\geq  \frac{\sigma^2 - \sigma_{\vv}^2 - \eta d_2^2}{d_1^2 - d_2^2}
\end{align*}
Let us simplify the numerator $Z =\sigma^2 - \sigma_{\vv}^2 - \eta d_2^2$. Since we are truncating the sample, we have $(1 + c_3\eta \log^{2/\gamma}\frac{n}{\eta}) \sigmahat_{\vv}^2 \le \sigma^2$. Here we also assume that $\eta\le \frac{1}{C}$ for a sufficiently large $C$ so that $\frac{1}{1-c\eta}$ is less than some constant.
\begin{align*}
Z & \geq \sigma^2 - \frac{\sigma^2}{1+c_3\eta \log^{2/\gamma} \frac{n}{\eta}}\pr{\frac{1}{1-c\eta} + \eta  \frac{c_2^2 \log^{2/\gamma} \frac{n}{\eta}}{4}}\\
&\ge \sigma^2 -\sigma^2\pr{\frac{1 - c\eta + (c\eta)^2 + \eta  \frac{c_2^2 \log^{2/\gamma} \frac{n}{\eta}}{4}}{1 + c_3\eta \log^{2/\gamma} \frac{n}{\eta}}} \\
&\ge \sigma^2 \pr{\frac{c_3\eta \log^{2/\gamma} \frac{n}{\eta} - (c\eta)^2 - \eta \frac{c_2^2 \log^{2/\gamma} \frac{n}{\eta}}{4} }{1 + c_3 \eta \log^{2/\gamma}\frac{n}{\eta}}}
\end{align*}
Recall that $\sigma^2 \ge (1-\eta) \|\SSigma\|_2$ by (\ref{eq:sigmanorm}). Then as long as $c_3$ is a sufficiently large constant, we have
\begin{align*}
Z &\ge \frac{\|\SSigma\|_2}{4} \pr{\frac{c_3\eta \log^{2/\gamma} \frac{n}{\eta} }{1 + c_3 \eta \log^{2/\gamma}\frac{n}{\eta}}}
\end{align*}
Then combining $Z$ with the denominator from earlier and using the fact that $d_1 \le c_1\sqrt{n\|\SSigma\|_2}\log^{1/\gamma}\frac{n}{\eta}$, we get:
\begin{align*}
r &\ge \frac{\|\SSigma\|_2 \pr{\frac{c_3\eta \log^{2/\gamma} n /\eta}{1 + c_3 \eta \log^{2/\gamma}\frac{n}{\eta}}}}{4c_1^2 \|\SSigma\|_2 n \log^{2/\gamma} \frac{n}{\eta} }\\
&\ge \frac{c_3\eta  }{4c_1^2 n (1 + c_3 \eta \log^{2/\gamma}\frac{n}{\eta})}
\end{align*}
Then $r \ge O\pr{\min\left\{\frac{\eta}{n}, \frac{1}{n \log^{2/\gamma} \frac{n}{\eta}}\right\}}$, so the algorithm will terminate in a nearly linear number of iterations.
\qed



\section*{Open Questions}
An immediate open question is whether the our analysis of the mean estimation algorithm is tight and the $\sqrt{\log n}$ is avoidable. For special distributions including Gaussians, \cite{DKKL16} give an algorithm with higher sample complexity and error $\eta \sqrt{\log \frac{1}{\eta}}$ rather than $\eta \sqrt{\log n}$ or $ \sqrt{\eta \log n}$ as in Theorem~\ref{thm:mainGauss}. An open question is to give an $O(\eta)$ approximation. For the more general distributions considered here, the dependence on $\eta$ must grow as at least $\eta^{3/4}$; it is open to find an algorithm that achieves $O(\eta^{3/4})$ error (our guarantee for the general setting has error $O(\sqrt{\eta \log n})$).  Other open problems include agnostic learning of a mixture of two arbitrary Gaussians and agnostic sparse recovery. 

\section*{Acknowledgment}

We thank Chao Gao and Roman Vershynin for helpful discussions. We would also like to thank the anonymous reviewers for useful suggestions. This research was supported in part by NSF awards CCF-1217793 and EAGER-1555447.

\bibliographystyle{alpha}
\bibliography{ref,spectralbook,ICA_bibliography} 
 
\end{document}